\documentclass[a4paper,fleqn,useAMS,usenatbib]{mnras}

\usepackage{graphicx,amsmath,amssymb,multirow,units}
\usepackage{lscape}
\usepackage{afterpage}
\usepackage{color}
\usepackage[T1]{fontenc}
\usepackage{ae,aecompl}
\usepackage{hyperref}

\newcommand{\msun}{\mbox{\,$\rm M_{\odot}$}}        
\newcommand{\lsun}{\mbox{\,$\rm L_{\odot}$}}        
 
\newcommand{\lcoa}{\mbox{\,$ L^{\prime}_{10}$}}
\newcommand{\lcob}{\mbox{\,$ L^{\prime}_{21}$}}
\newcommand{\Md}{\mbox{\,$M_{\rm d}$}}
\newcommand{\Ms}{\mbox{\,$M_{\ast}$}}
\newcommand{\Mh}{\mbox{\,$\rm M_{H2}$}}
\newcommand{\Lir}{\mbox{\,$\rm L_{IR}$}}
\newcommand{\Lco}{\mbox{\,$\rm L^{\prime}_{CO}$}}
\newcommand{\COa}{\mbox{\,$\rm{^{12}CO(1-0)}$}}
\newcommand{\COb}{\mbox{\,$\rm{^{12}CO(2-1)}$}}
\newcommand{\COc}{\mbox{\,$\rm{^{12}CO(3-2)}$}}
\newcommand{\asec}{\ensuremath{^{\prime\prime}\,}}

\newcommand{\kms}{\mbox{\,$\rm{km\,s^{-1}}$}}
\newcommand{\tA}{\mbox{\,$\rm{T_A^{\ast}}$}}
\newcommand{\tmb}{\mbox{\,$\rm{T_{mb}}$}}

\newcommand{\tba}{\mbox{\,$\rm{T^{10}_{B}}$}}
\newcommand{\tbb}{\mbox{\,$\rm{T^{21}_{B}}$}}
\newcommand{\aco}{\mbox{\,$\alpha_{\rm{CO}}$}}
\newcommand{\mic}{$\mu $m}

\title[BADGRS: A most unusual ISM.]{The unusual ISM in Blue and Dusty Gas Rich Galaxies (BADGRS).}

\author[L. Dunne et al.]{
L. Dunne,$^{1,2}$\thanks{E-mail:ldunne@roe.ac.uk} 
Z. Zhang,$^{2,3}$
P. De Vis,$^{4}$
C.~J.~R. Clark,$^{1}$
I. Oteo,$^{2,3}$
S.~J. Maddox,$^{1,2}$
\and
P. Cigan,$^1$
G. de Zotti,$^5$
H.~L. Gomez,$^1$
R.~J. Ivison,$^{2,3}$
K. Rowlands,$^6$
\and
M.~W.~L. Smith,$^{1}$
P. van der Werf,$^7$
C. Vlahakis,$^8$
J.~S. Millard$^1$
\\
$^{1}$School of Physics \&\ Astronomy, Cardiff University, Queens Buildings, The Parade, Cardiff, CF24 3AA, UK \\
$^{2}$SUPA, Institute for Astronomy, University of Edinbugh, Royal Observatory, Blackford Hill, Edinbugh EH9 3HJ, UK\\
$^{3}$ESO, Karl-Schwarzschild-Strasse 2, 85748 Garching, Germany\\
$^{4}$Institut d'Astrophysique Spatiale, CNRS, Universit\'{e} Paris-Sud, Universit\'{e} Paris-Saclay, Bat. 121, 91405, Orsay Cedex, France\\
$^5$INAF-Osservatorio Astronomico di Padova, Vicolo Osservatorio 5, I-35122, Padova, Italy\\
$^{6}$Johns Hopkins University, Bloomberg Center, 3400 N. Charles St, Baltimore, MD 21218, USA\\
$^{7}$Leiden Observatory, Leiden University, P.O. Box 9513, NL -2300 RA Leiden \\
$^8$NRAO, 520 Edgemont Road, Charlottesville, VA 22903-2475\\
}

\date{\vspace*{-5em}\today}

\pubyear{2017}

\begin{document}
\label{firstpage}
\pagerange{\pageref{firstpage}--\pageref{lastpage}} 
\maketitle

\begin{abstract}
The {\em Herschel}-ATLAS unbiased survey of cold dust in the local
Universe is dominated by a surprising population of very blue
($FUV-K<3.5$), dust-rich galaxies with high gas fractions
($\rm{f_{HI}=M_{HI}/(\Ms+M_{HI})}>0.5$). Dubbed `Blue and Dusty Gas
Rich Sources' (BADGRS) they have cold diffuse dust temperatures, and
the highest dust-to-stellar mass ratios of any galaxies in the local
Universe. Here, we explore the molecular ISM in a representative
sample of BADGRS, using very deep $\rm{CO(J_{up}=1,2,3)}$
observations across the central and outer disk regions. We find very
low CO brightnesses ($T_p = 5-30$ mK), despite the bright far-infrared
emission and metallicities in the range $0.5 < Z/Z_{\odot} < 1.0$. The
CO line ratios indicate a range of conditions with
$R_{21}=\rm{T_b^{21}/T_b^{10}=0.6-2.1}$ and
$R_{31}=\rm{T_b^{32}/T_b^{10}=0.2-1.2}$. Using a metallicity
dependent conversion from CO luminosity to molecular gas mass we find
$\Mh/\Md\sim 7-27$ and $\Sigma_{\rm{H2}} = 0.5-6\,\msun\,\rm{pc^{-2}}$,
around an order of magnitude lower than expected. The BADGRS have
lower molecular gas depletion timescales ($\tau_d\sim 0.5$ Gyr) than
other local spirals, lying offset from the Kennicutt-Schmidt relation
by a similar factor to Blue Compact Dwarf galaxies. The cold diffuse
dust temperature in BADGRS (13--16 K) requires an interstellar
radiation field 10--20 times lower than that inferred from their
observed surface brightness.  We speculate that the dust in these
sources has either a very clumpy geometry or a very different opacity
in order to explain the cold temperatures and lack of CO
emission. BADGRS also have low UV attenuation for their UV colour
suggestive of an SMC-type dust attenuation curve, different star
formation histories or different dust/star geometry. They lie in a
similar part of the IRX-$\beta$ space as $z\sim 5$ galaxies and may be
useful as local analogues for high gas fraction galaxies in the early
Universe.
\end{abstract}


\section{Introduction}
\label{IntroS}
Blind surveys in unexplored wave-bands often reveal new insights into
the process of galaxy evolution, by virtue of a different set of
selection characteristics
\citep{Schmidt1963,deJong1984,Meegan1992,SIB1997,Eales2018}. The
Herschel Astrophysical Terahertz Large Area Survey (H-ATLAS:
\citealt{Eales2010}) is the {\em first blind survey} of the local
Universe at sub-millimetre wavelengths (250\mic). In contrast to {\em
  IRAS}, which was only sensitive to the $\sim 10$ percent (by mass)
of dust heated strongly enough to radiate substantially at 60\mic,
H-ATLAS selects galaxies based on the $\sim 90$ percent (by mass) of
dust heated to 15--25 K by the diffuse interstellar radiation field
\citep{Helou1986,Dunne2000,Draine2007}.

\citet{Clark2015} used H-ATLAS to produce the first local volume
limited sample selected on the basis of cool dust emission (HAPLESS:
$z<0.01$) and showed that the 250\mic\ selection probed a uniform range
in gas fraction from 10--90\%, in stark contrast to optically selected
samples, which are comprised mainly of galaxies dominated by their
stellar component. A new population of `Blue and Dusty Gas Rich
Sources' (henceforth BADGRS) were identified, comprising more than 50
percent of the 250-\mic\ selected galaxies in HAPLESS. The properties
of these BADGRS detailed in \citet{Clark2015} are: very blue UV-NIR
colour ($\rm{FUV-K}<3.5$), intermediate stellar mass ($10^{8} < \Ms <
10^{10}$ \msun), flocculent or irregular morphologies and high gas
fractions: $\rm{f_{HI}=M_{HI}/(\Ms+M_{HI}})>0.5$.\footnote{See
  \citet{Clark2015,deVis2017} for details of the H{\sc i} data.}  For comparison, the
average gas fraction for the K-band selected Herschel Reference Survey (HRS:
\citealt{Boselli2010}), is $\sim$18 percent.

Low mass, blue and gas-rich galaxies are commonly perceived to be low
in dust content \citep{Hunter1989,Giovanelli1995,Tully1998}, however
the BADGRS in H-ATLAS have the highest dust-to-stellar mass ratios of
any population in the local Universe
\citep{Clark2015}. \citet{deVis2017} found that the dust-to-stellar
mass ratio peaks at a gas fraction $\sim 70$ percent, with {\em BADGRS
  locating the peak dust content relative to stars and relative to
  baryons} (however, not relative to gas as this is a monotonically
increasing quantity). As these galaxies have only moderate infra-red
luminosities ($\rm 8.0< Log\,\Lir<10.0$), and often low ratios of
$S_{60}/S_{100}$, they are mostly undetected or very faint in the {\em
  IRAS} bands. A wide area, sensitive sub-millimetre survey such as
H-ATLAS was therefore required to identify this population. In
addition to making up more than half the number density of submm
galaxies in the local volume, BADGRS account for 30\% of the
integrated dust mass density and 20\% of the star formation rate
density in the volume probed by HAPLESS, yet contain only 6\% of the
stellar mass \citep{Clark2015}.

\citet{deVis2017CE} used chemical and dust evolution modelling to show
that different processes must be acting in either the formation or
destruction of dust at the same gas fraction in order to explain the
differences between BADGRS and the more commonly studied dust-poor,
metal-poor, gas-rich galaxies in the Herschel studies of H{\sc
  i}-selected (HIGH: \citealt{deVis2017}) and dwarf galaxies (DGS:
\citealt{Madden2013}). BADGRS may be therefore be a Rosetta Stone for
understanding the changes in both the dust and gas properties of the
ISM when it becomes enriched with dust, yet is still dominated by gas
rather than stars (a situation which will be more common in the early
universe).

\begin{figure*}
\includegraphics[scale=0.29]{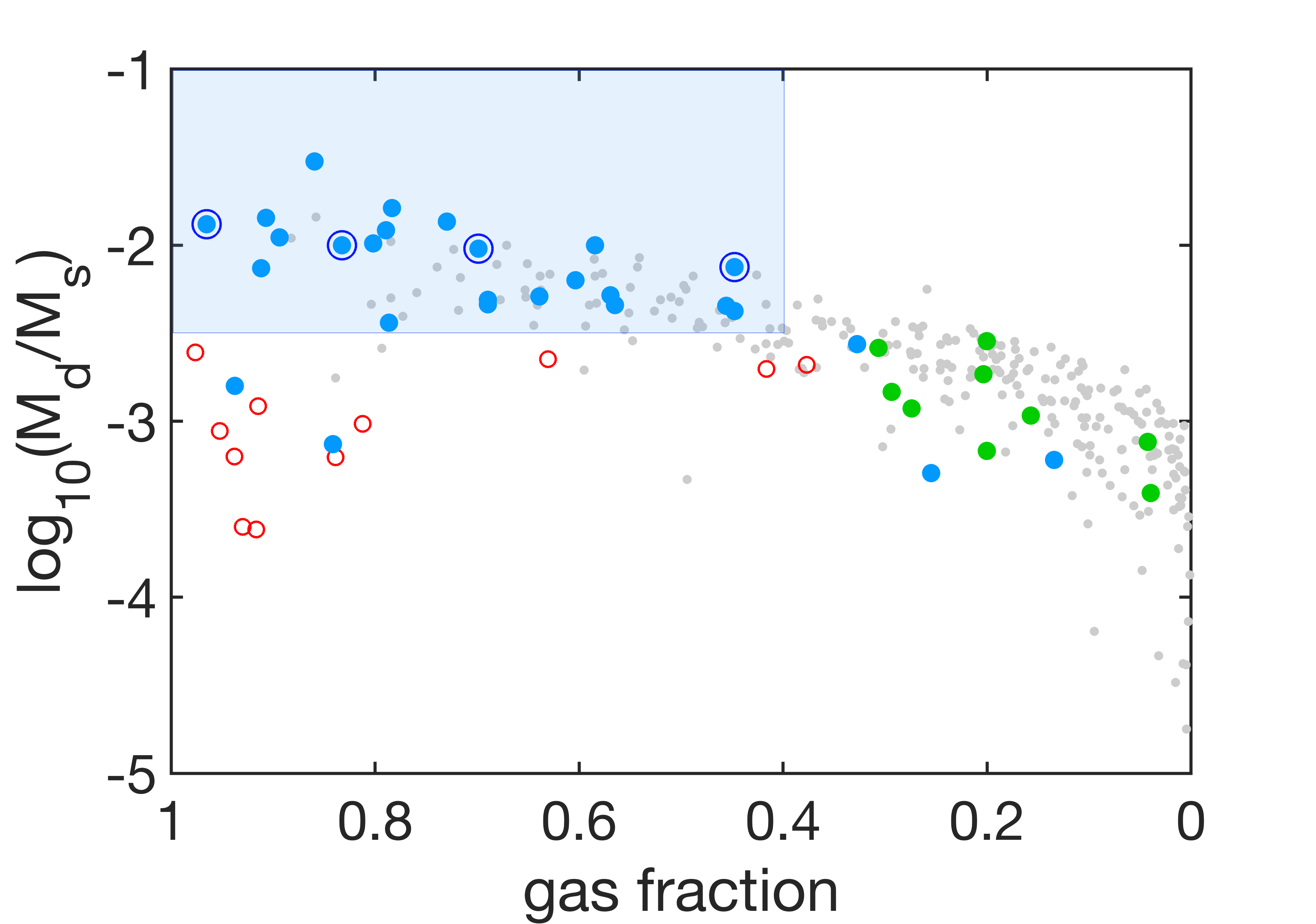}
\includegraphics[scale=0.30]{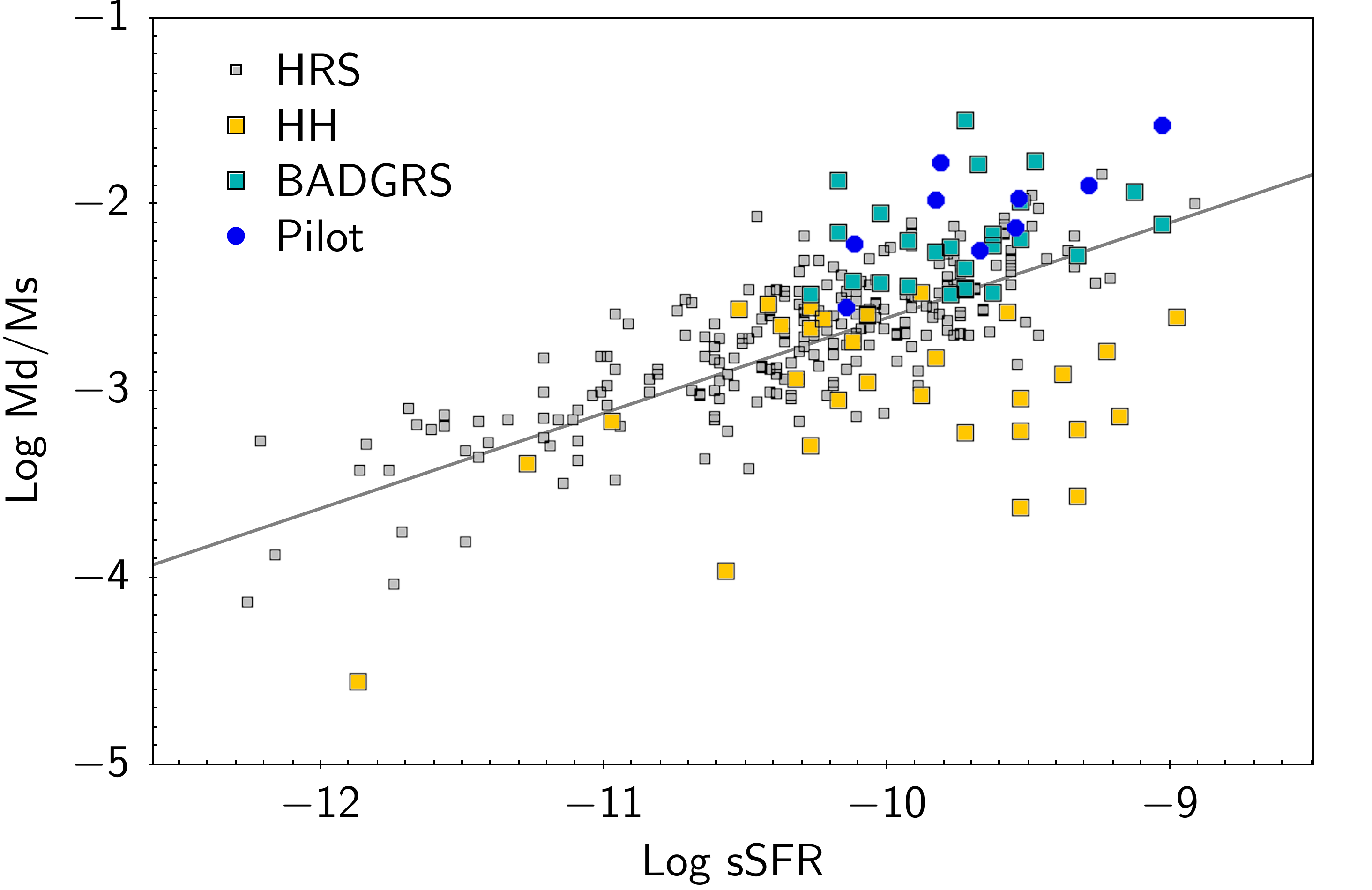}
\caption{\label{sampleF} {\bf Left}: Specific dust mass (\Md/\Ms) vs
  gas fraction for local galaxy samples. Galaxies with $\rm FUV-K <
  3.5$ and which are detected in H-ATLAS are denoted by blue
  circles. The parent samples are the dust selected local volume
  sources from \citet{Clark2015} and the H{\sc i} selected sample from
  \citet{deVis2017}. Dust and H{\sc i} selected sources from the
  parent samples which do not have both the blue colour and H-ATLAS
  detections are shown as filled green and red open circles
  respectively. The shaded box indicates our definition of BADGRS as
  described in the text. Our CO BADGRS sample is highlighted by dark
  blue circles surrounding the points. K-band selected sources (grey)
  are from HRS \citep{Boselli2010}. Dust masses are all scaled to
  $\kappa_{250}=0.56\,\rm{m^2\,kg^{-1}}$ and are determined in an
  equivalent way, using either {\sc magphys} or two temperature MBB
  fitting. For more information on dust mass determination, see
  Section~\ref{DustS}. {\bf Right}: Dust mass versus star formation
  rate (normalised by stellar mass). BADGRS have more dust per unit
  SFR than typical galaxies and the wide spread of dust per SFR is
  apparent at high sSFR due to the inclusion of the H{\sc i} selected,
  dust poor sources from HIGH \citep{deVis2017}. BADGRS from both HIGH
  and HAPLESS are shown in cyan while the individual regions of the
  pilot sample presented here are blue circles. Gold squares are the
  HIGH and HAPLESS sources which do not meet the BADGRS criteria of
  being blue {\em and} dusty.}
\end{figure*}

Figure~\ref{sampleF}(a) shows the dust content relative to stellar
mass versus the gas fraction for sources from the dust (HAPLESS),
H{\sc i} (HIGH) and K-band (HRS) selected samples
\citep{Clark2015,deVis2017,Boselli2010}. Blue circles represent those
HAPLESS and HIGH sources with $\rm FUV-K<3.5$ that are in the H-ATLAS
catalogue. The remaining HAPLESS sources with redder colours are shown
as green circles, and HIGH sources that are not in the H-ATLAS
catalogues are red open circles. The K-band selected HRS galaxies are
shown in grey; the nature of their K-band selection ensures that most
are evolved and gas poor (residing at the right of the plot). The $\rm
FUV-K<3.5$ dust detected sources (blue circles) are mostly located in
a region in the upper left quadrant of this parameter space -
i.e. they are both dust and gas rich. The few exceptions are early
type galaxies at lower gas fraction undergoing interactions with gas
rich galaxies, and two low dust mass sources at high gas fraction
which are the very local dwarf galaxies UM451 and UM452. We now make
the working definition of `Blue and Dusty Gas Rich Sources' (BADGRS)
to be ($FUV-K<3.5$) {\em and} ($\rm{Log\,\Md/\Ms > -2.46 -
  log(\kappa_{250}/0.56)}$), this consists of the blue points
contained within the shaded box in Figure~\ref{sampleF}(a).
Figure~\ref{sampleF}(b) shows the relation between dust mass and star
formation rate, (normalised by stellar mass to remove the tendency for
large galaxies to have more of everything). This correlation has been
found to hold for local galaxies and SMGs at high redshift
\citep{daCunha2010,DJBSmith2012,Rowlands2014}. BADGRS from the HIGH
and HAPLESS samples are denoted as cyan squares, while the individual
regions for our four pilot targets are shown as blue circles. BADGRS
have more dust per unit SFR than similarly high sSFR galaxies which do
not meet the colour and dust content threshold.

BADGRS possess common properties which are intriguing and indicate an unusual ISM:
\begin{enumerate}
{\item {\em Hot but cold:} Their diffuse ISM dust temperatures are far colder (12--16K)
  than the average for spirals and dwarfs (18--32K), despite their
  relatively intense UV emission \citep{Clark2015}.}  
{\item {\em Blue but dusty:} They have
  very little UV obscuration compared to other galaxies with similar
  dust masses, or similar UV colours (\citealt{deVis2017}, Dunne et
  al. {\em in prep}).}  
{\item {\em Metal rich but CO poor:} We found that their predicted CO
  luminosities were far lower than expected given their dust emission,
  assuming they lay on the \Ms-\Mh\ scaling relations of local galaxies
  \citep[e.g.][]{Saintonge2011,Bothwell2014}.}
\end{enumerate}

It was this final puzzle which led us to propose extremely sensitive
CO observations with the IRAM 30-m telescope in order to test whether
these galaxies were indeed unusually rich in dust and IR emission
relative to their molecular gas content. This first paper motivates
the case for an unusual ISM in BADGRS by presenting our first complete
data-set of \COa, \COb\ and \COc\ measurements in 9 regions across four
representative examples of BADGRS from \citet{Clark2015}. We compare
the CO properties to those of the dust (as traced by {\em Herschel})
in the same regions. We will also investigate the dust heating and obscuration puzzles
and their potential relationship to the CO observations.

In Sections \ref{SampleS} and ~\ref{COdataS} we describe the sample
and CO observations. In Section~\ref{COpropsS} we discuss the CO and
molecular gas properties. In Section~\ref{DustS} we describe the dust
properties and compare those of other samples from the literature. In
Section~\ref{DiscS} we discuss the ISM of BADGRS and look for possible
explanations for their unusual properties. Throughout we use a cosmology with
$\Omega_m = 0.27,\,\Omega_{\Lambda} =0.73$ and $H_o = 71\,
\rm{km\,s^{-1}\,Mpc^{-1}}$.



\section{Sample and Data}
\subsection{The pilot sample of BADGRS}
\label{SampleS}

To undertake an initial detailed study of BADGRS we selected a small
pilot sample which was representative of the group. The very local
($D<45$ Mpc) nature of the sources in \citet{Clark2015} means they
typically have large angular sizes making a detailed study of the
dust, gas and optical properties in a large sample prohibitive.  We
chose four galaxies which spanned the range in stellar mass, gas
fraction, colour (FUV-K) and morphology seen across the BADGRS
population and which were bright enough and resolved enough to provide
excellent targets for follow up. Figure~\ref{sampleF}(a) shows our pilot
sample as blue circles surrounding blue dots and Table~\ref{sampleT}
summarises their global properties.

The FIR data-set used for the {\em Herschel}-ATLAS local volume sample
and this paper is the H-ATLAS DR1 release described in
\citet{Valiante2016} and \citet{Bourne2016}. It consists of imaging in
five bands from 100--500\mic, covering the 161 sq deg of H-ATLAS which
coincides with the equatorial fields of the Galaxy And Mass Assembly
(GAMA; \citealt{Driver2011}) survey. The FUV-22-\mic\ imaging was
provided by GAMA \citep{Driver2016}, who consolidated the data from
multiple public surveys into a format ideal for multi-band
matched-aperture photometry. The photometry was performed as described
in \citet{Clark2015} and \citet{deVis2017}. Optical spectra of the
central region, and some H{\sc ii} regions were taken from GAMA and
SDSS \citep{Hopkins2013,SDSSdr10} and used to provide redshifts and
estimate metallicities. Composite colour images of the galaxies are
shown in Figure~\ref{spectraF}, their semi-major axis size ranges from
1.5\arcmin to 5\arcmin, and they have inclinations in the range
$50-80\deg$.

The galaxies were observed using 9 pointings with the IRAM 30-m and
APEX\footnote{This publication is based on data acquired with the
  Atacama Pathfinder Experiment (APEX). APEX is a collaboration
  between the Max-Planck-Institut fur Radioastronomie, the European
  Southern Observatory, and the Onsala Space Observatory.}  to probe
the properties of the molecular gas across the central and disk
regions. The regions sampled are shown in Figure~\ref{spectraF} and
the positions are listed in Table~\ref{COdataT}. The positions are
named such that the central position is `C' and the offset positions
are `O1,O2'. At the distance of the galaxies, the 22\asec beam of the
IRAM 30-m at 115-GHz corresponds to 2-3 kpc linear size.

\subsubsection*{Atomic gas properties}

The very blue $\rm FUV-K<3.5$ sources in the HAPLESS \citep{Clark2015}
sample have a high average $f_{\rm HI}=0.66$, due to the correlation
between $\rm FUV-K$ colour, sSFR and gas content. Our additional dust
content selection criteria (Fig~\ref{sampleF}a) makes the BADGRS
sample exclusively gas rich. The global atomic gas fractions for the pilot sample
are listed in Table~\ref{sampleT}. The H{\sc i} measurements
(presented in \citealt{Clark2015,deVis2017}) are from single dish
surveys with the Parkes (HIPASS; \citealt{Meyer2004}) and Arecibo
(ALFALFA; \citealt{Haynes2011}) telescopes and as such have poor
angular resolution ranging from 5-15\arcmin. This means that apart
from a global measure of their atomic gas content, we can do no more
at this time to explore the relationship between the dust, molecular
and atomic gas content as the dust and molecular gas are observed with
2--3 kpc angular resolution in the galaxy disks while a large fraction
of the H{\sc i} could reside at much larger radii, as is common for
lower surface brightness blue galaxies such as these. A future study
(Dunne et al. in prep) will investigate the atomic gas properties in
detail using higher resolution interferometric H{\sc i} data.

\subsubsection*{Metallicities}

There are local metallicity estimates for all of our gas and dust
comparisons, as for each CO pointing there is at least one metallicity
measure available within the 115-GHz beam area of the IRAM 30-m.
For NGC~5496, UGC~9215 and UGC~9299 metallicities were taken from
\citet{deVis2017CE}, who used the SDSS and GAMA spectra and strong
line ratios, and are quoted in the \citet{Pettini2004} O3N2
calibration. There are 10 measurements in total across the 6 CO
pointings, as several H{\sc ii} regions were observed in some
galaxies.  For NGC~5584, we use the comprehensive set of measurements
as part of the HST Cepheid study by \citet{Riess2011} which results in
14 metallicity measures across the three pointings. We convert the
metallicity calibration used in \citet{Riess2011} to the O3N2 relation
using the conversions given by \citet{Kewley2008}. The local
metallicity measures match the recent determination of the
$\Sigma_{\ast}-Z$ relation using the MaNGA survey
\citep{BarreraB2016}. Metallicities for each region are listed in
Table~\ref{DustT} and range from $\rm{12+log(O/H)= 8.3-8.7}$ ($0.4-1\, Z_{\odot}$), with
an average of 8.51 ($0.66\,Z_{\odot}$), assuming $\rm{12+log(O/H)_{\odot}=8.67}$ \citep{Asplund2009}.

\subsubsection*{Global properties}
Stellar masses, infrared luminosities and global FUV attenuations are
taken from \citet{deVis2017} who used {\sc magphys}, an energy balance
SED fitting code \citep{daCunha2008}, and 21-band matched aperture
photometry measurements from FUV-500$\mu$m. We estimate the recent SFR
using the relation \citep{Hao2011,Kennicutt2012}
\begin{equation}
\rm{Log\,SFR = Log(L_{FUV}+\eta\, \Lir) - 43.35} \label{SFRE}
\end{equation}
with luminosities in units of erg $\rm{s}^{-1}$. The parameter $\eta$
describes the contribution of dust heated by young stars to the \Lir\
and we take the value of $\eta=0.46$ from \citet{Hao2011}. Comparing
the total SFR to that traced by the uncorrected FUV gives us
unobscured star formation rate fractions of 53, 64, 58, and 81 percent
for the four sources respectively\footnote{The {\sc magphys} SFR PDFs
  for these very blue galaxies were often multi-peaked meaning the
  median value was not a good representation of the most likely SFR
  value. See \citet{deVis2017,Schofield2017} for details.}

We have used the original two component modified blackbody method of
deriving the dust properties, as presented in \citet{Clark2015} and
described originally in \citet{Dunne2001} because {\sc magphys}
consistently underpredicts the 500\mic\ flux for all the sources which
show cold dust temperatures in \citet{Clark2015}. We believe this is
due to some of the complex priors used in the energy balance and
infra-red SED construction in {\sc magphys}. More details of the dust
SED fitting are given in Section~\ref{DustS}.

These global properties are listed in Table~\ref{sampleT}.


\begin{table*}
\caption{\label{sampleT} Properties of the four BADGRS in our pilot sample.}  \centering
\begin{tabular}{lcccccccccc}
\hline
Name & $v_{\rm{lsr}}$ & D & Log \Ms & Log $\mu_{\ast}$ & Log \Lir & Log \Md & Log $M_{\rm{HI}}$ & FUV-K  & sSFR & $f_{\rm{HI}}$ \\
     & (\kms) & (Mpc) &   (\msun)  & (\msun\ $\rm{kpc^{-2}}$)  & (\lsun)  & (\msun)  & (\msun) &  & log (yr$^{-1}$)  &      \\
\hline
NGC~5584 & 1638 & 30.2 & 9.98 & 7.02 & 10.04 & 7.80 & 9.76 & 2.68  & $-9.71$ & 0.44 \\
NGC~5496 & 1541 & 27.4 & 9.46 & 6.64 & 9.51 & 7.53 & 10.03 & 2.34 & $-9.62$ & 0.83 \\
UGC~9215 & 1397 & 25.6 & 9.31 & 6.93 & 9.57 & 7.28 & 9.56 & 2.06  & $-9.47$ & 0.70\\
UGC~9299 & 1539 & 28.3 & 8.61 & 6.67 & 8.82 & 6.74 & 9.94 & 0.93  & $-9.20$ & 0.97\\
\hline
\end{tabular}
\flushleft{\small{$v_{\rm{lsr}}$ is the recessional velocity. Distance
    is taken from \citet{deVis2017} and is local flow corrected
    following \citet{Baldry2012}, \Ms\ and \Lir\ from MAGPHYS
    \citep{deVis2017}, \Md\ from 2-temperature MBB fit, $M_{\rm HI}$
    for NGC~5584 and NGC~5496 from HIPASS \citep{Meyer2004} and for
    UGC~9215 and UGC~9299 from ALFALFA \citep{Haynes2011}, SFR is from
    UV+TIR following Eqn~\ref{SFRE}. $\rm{f_{HI}=M_{HI}/(M_{\ast}+M_{HI}})$.
    $\rm{M_{HI}}$ includes a factor 1.36 to account for He.}}
\end{table*}

\subsection{CO observations}
\label{COdataS}

Observations of the \COa\ and \COb\ lines were made with the IRAM 30-m
telescope between July 2 - July 4 2014. The EMIR spectrometer was
used, combining the E090 and E230 dichroics which allowed simultaneous
observations of both ${\rm ^{12}CO}$ lines, and additionally the
$\rm{^{13}CO(1-0)}$ line. The beam sizes are 21.5\asec and 10.7\asec
respectively. The WILMA back-end was used, producing a frequency
resolution of 0.51 MHz which was then hanning smoothed in data
analysis to typically 8--16 \kms. Wobbler switching with a throw of
240\asec, large enough to be off the target galaxy, was used. The
opacity at 225 GHz ranged from 0.09--0.6 over the three nights with
p.w.v. of 1.5--3mm. Each scan lasted 4.8 min, producing total integration
times of 57--211 mins per pointing. The spectra were reduced using the
{\sc gildas--class} software (Pety
2005)\footnote{http://www.iram.fr/IRAMFR/GILDAS/}, noisy channels were
flagged and replaced with the local noise from the channels either
side and a first-order baseline was subtracted from
each individual scan. Scans were then averaged using a
$t/T_{\rm{sys}}^2$ weighting and smoothed to the desired velocity
resolution. A linear baseline was then subtracted from the averaged
spectrum and the resultant velocity integrated line intensity (moment
0) was measured between the FWZI points. Units of intensity were
converted from the \tA\ scale to \tmb\ using the values in the IRAM 30-m
report (Kramer, Penalver \& Greve
2013)\footnote{http://www.iram.es/IRAMES/mainWiki/CalibrationPapers}
of $\rm{F_{eff}} = 0.94,\, 0.92$ and $\rm{B_{eff}} = 0.78,\, 0.59$ at
115, 230 GHz respectively. This gives conversions of $\tmb = 1.205\tA$
at 115 GHz and $\tmb = 1.559\tA$ at 230 GHz.

The line-widths (FWHM) were calculated as follows \citep{Heyer2001,Leroy2016} 
\begin{equation}
\Delta v_{CO} = 2.35 \frac{I}{\sqrt{2 \pi}I_p} \label{FWHME}
\end{equation}
where $I$ is the integrated intensity and $I_p$ the peak
intensity. This method is less prone to error than Gaussian fitting or
moment based methods.\footnote{Where the SNR was high enough and
  Gaussians could be fitted, we found excellent agreement between
  the fitted FWHM and the method above.} 
Where a line was not detected we report the 3$\sigma$ upper limit as 
\[
I_{CO} < 3\sigma\,\sqrt{\Delta v_{CO} \delta v}
\]
with $\sigma$ being the spectral r.m.s in mK, $\Delta v_{co}$ the
line-width expected and $\delta v$ the channel width in \kms.

Observations of the \COc\ line
were made with the APEX 12-m telescope between April and July 2015. The
beam-size at 352 GHz is 17.3\asec. The APEX-2 front-end was used with
the XFFTS back-end, resulting in an instrumental velocity resolution of
0.0665\kms, which was then hanning smoothed to 8--17\kms. The weather
was good with pwv ranging from 0.4-1.3 mm. On source integration times
ranged from 16--62 mins, resulting in typical rms of 1.4 mK in a 
30 \kms\ channel.

The spectra were reduced with {\sc class} in the same way as described
above for the IRAM 30-m, and the intensity units converted from \tA\ to \tmb\ using 
values of $\eta_f = 0.97$ and $\eta_{mb} = 0.73$
at 352 GHz, resulting in a conversion of $\tmb = 1.329\tA$

The integrated line intensities and other observational parameters are listed in Table~\ref{COdataT} and
the spectra are shown in Figure~\ref{spectraF}.

The line intensities were converted to fluxes using the appropriate
Jy/K conversion factors: $S_{\nu}$/\tmb = 5.0 Jy/K for \COa,
5.01 Jy/K for \COb\ and 30.85 Jy/K for \COc. The fluxes are listed in
Table~\ref{COpropsT}.

\begin{figure*}
\begin{minipage}[c]{\textwidth}
\begin{minipage}[c]{0.32\textwidth}
\includegraphics[width=\textwidth]{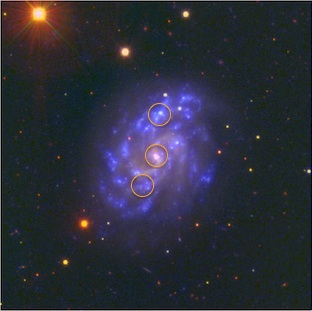}
\end{minipage}
\begin{minipage}[c]{0.32\textwidth}
\includegraphics[width=\textwidth]{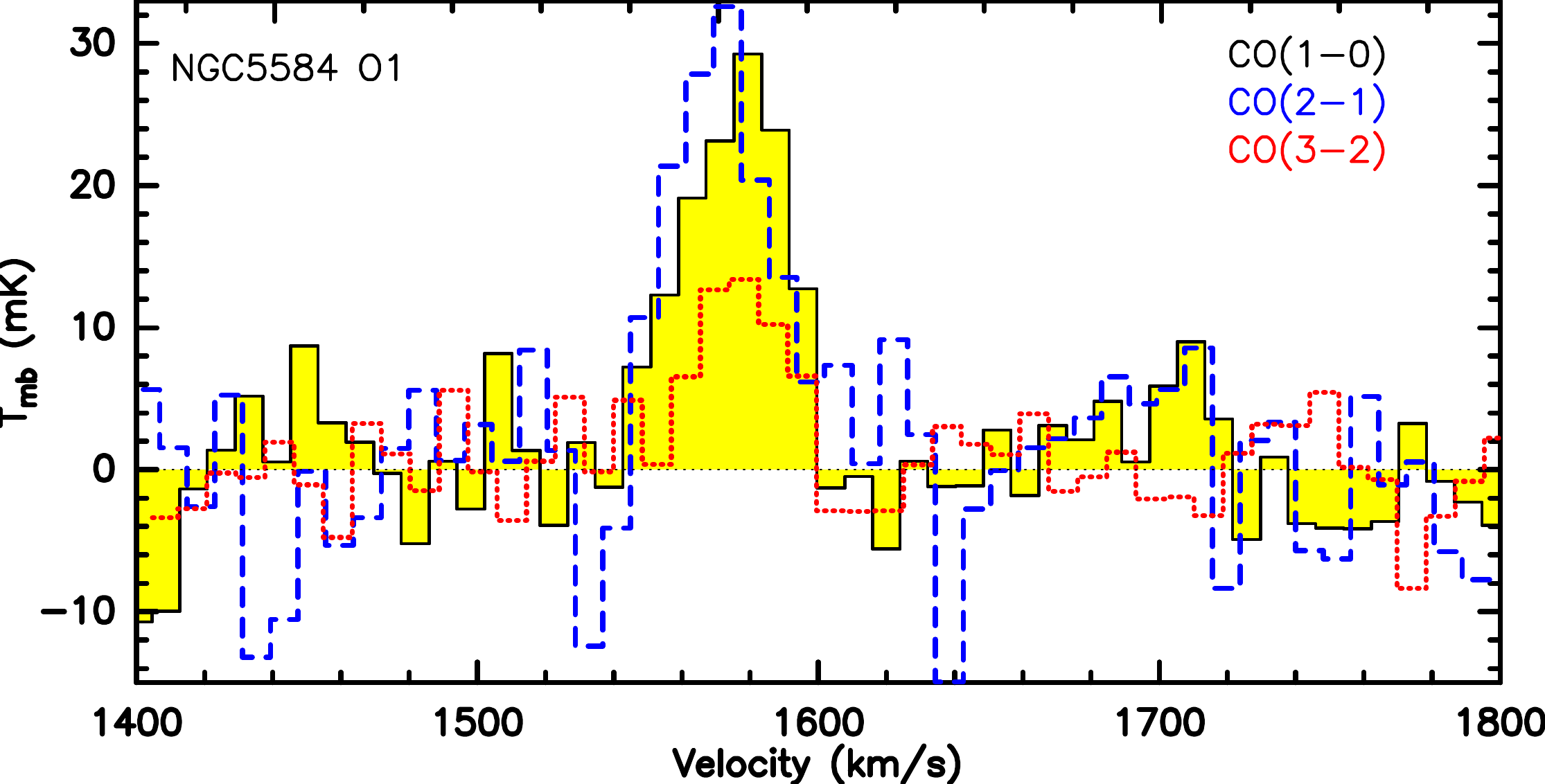}
\includegraphics[width=\textwidth]{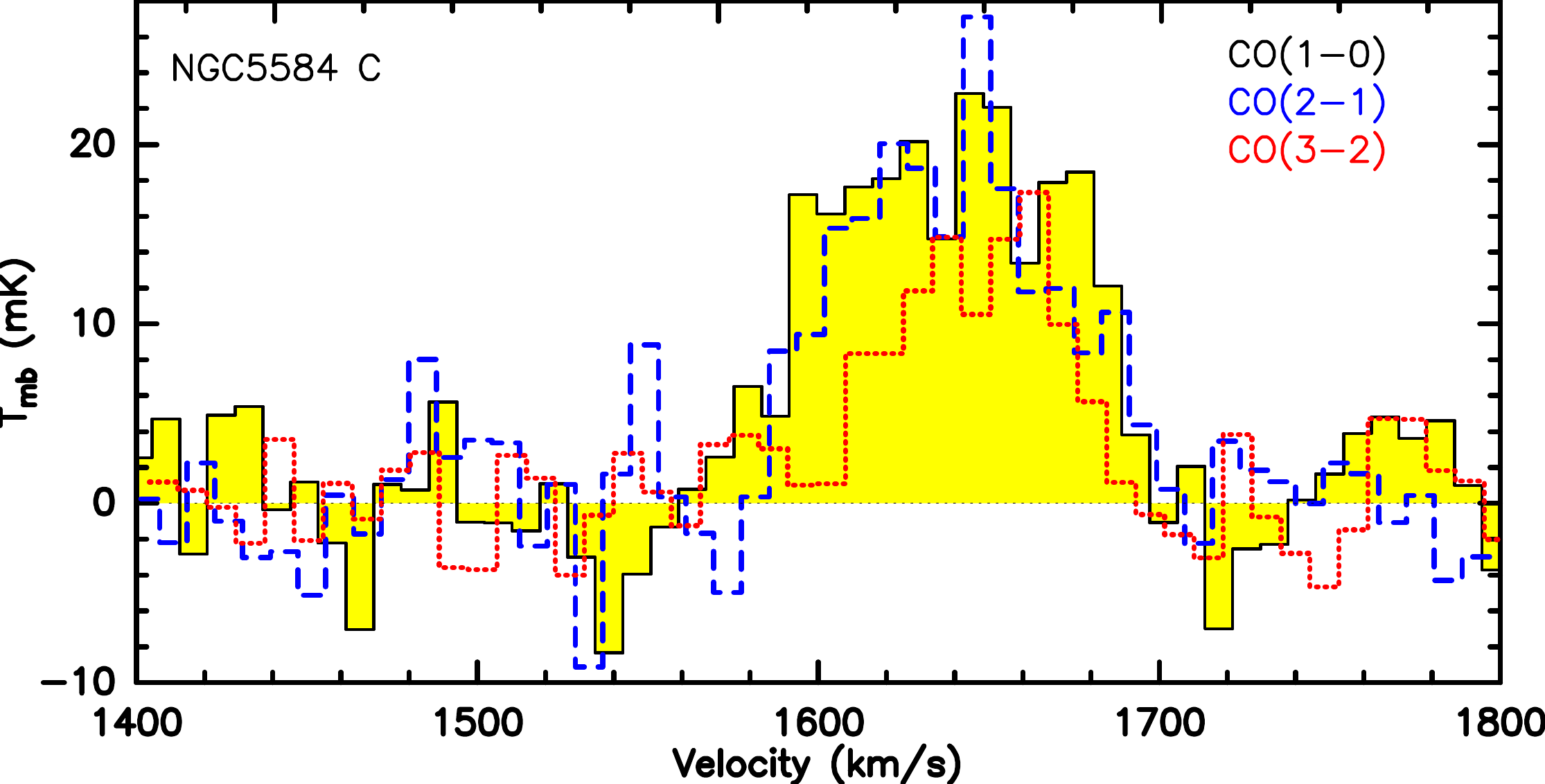}
\includegraphics[width=\textwidth]{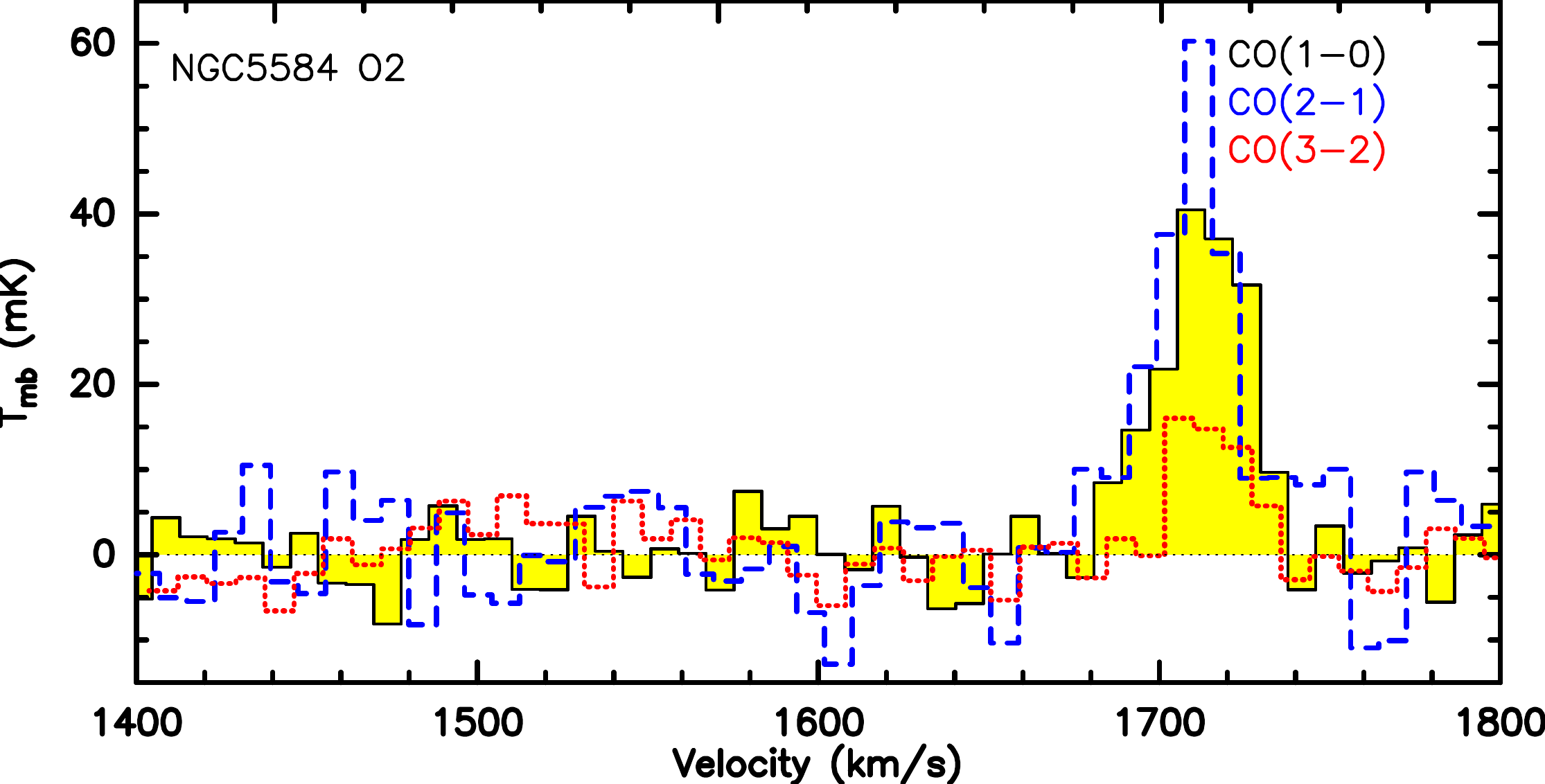}
\end{minipage}
\begin{minipage}[c]{0.32\textwidth}
\includegraphics[width=\textwidth]{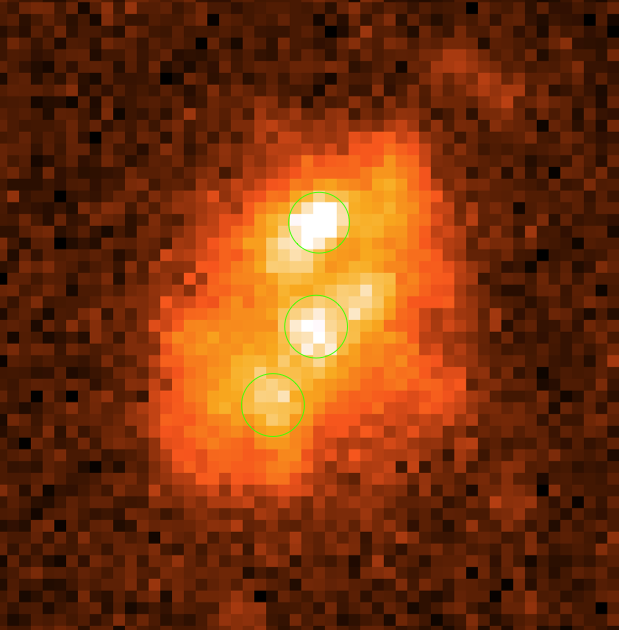}
\end{minipage}


 \begin{minipage}[c]{0.32\textwidth}
  \includegraphics[width=\textwidth]{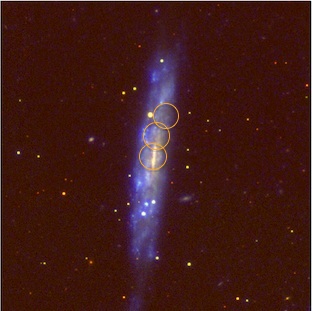}
 \end{minipage}
 \begin{minipage}[c]{0.32\textwidth}
 \includegraphics[width=\textwidth]{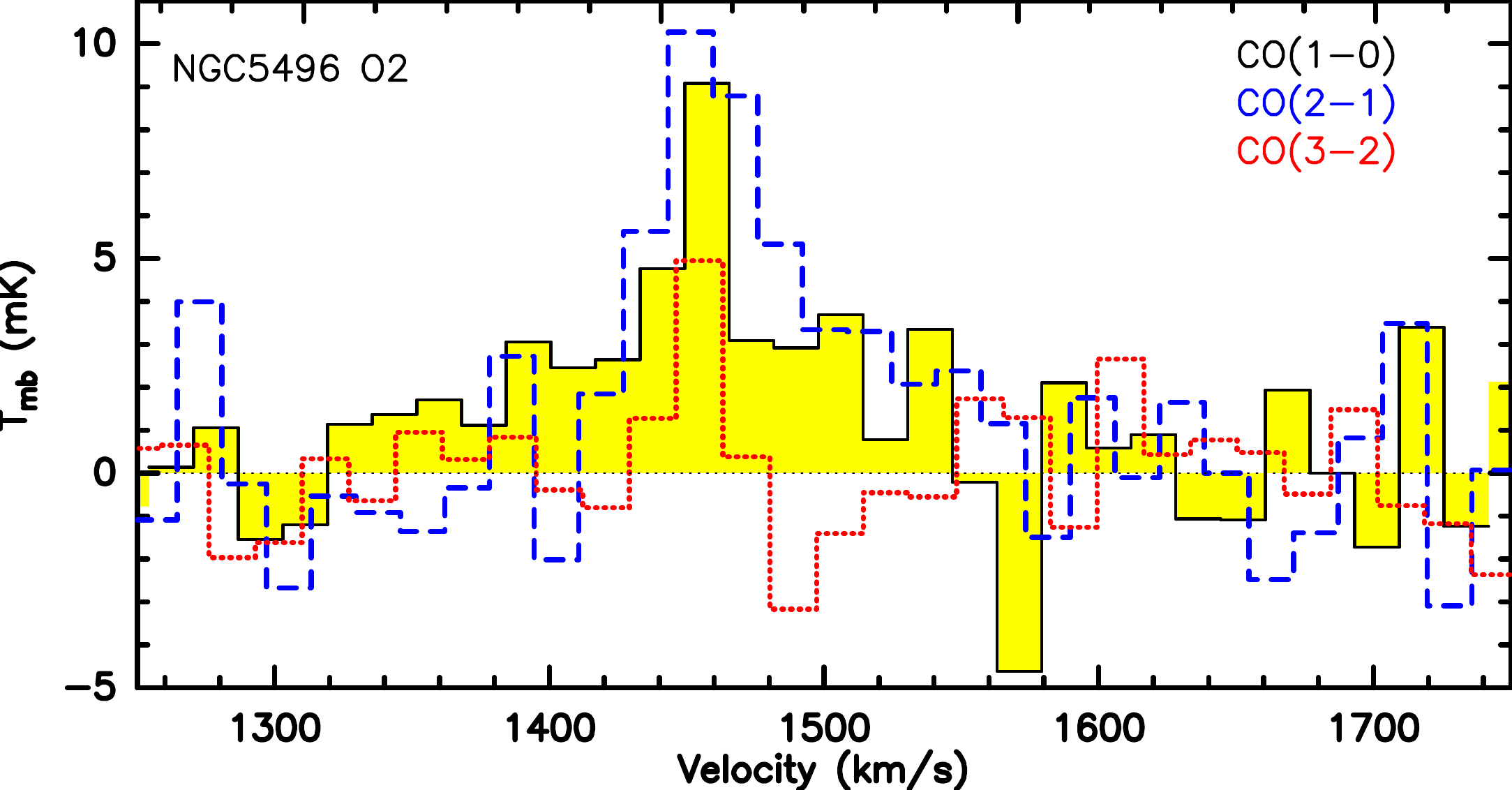}
 \includegraphics[width=\textwidth]{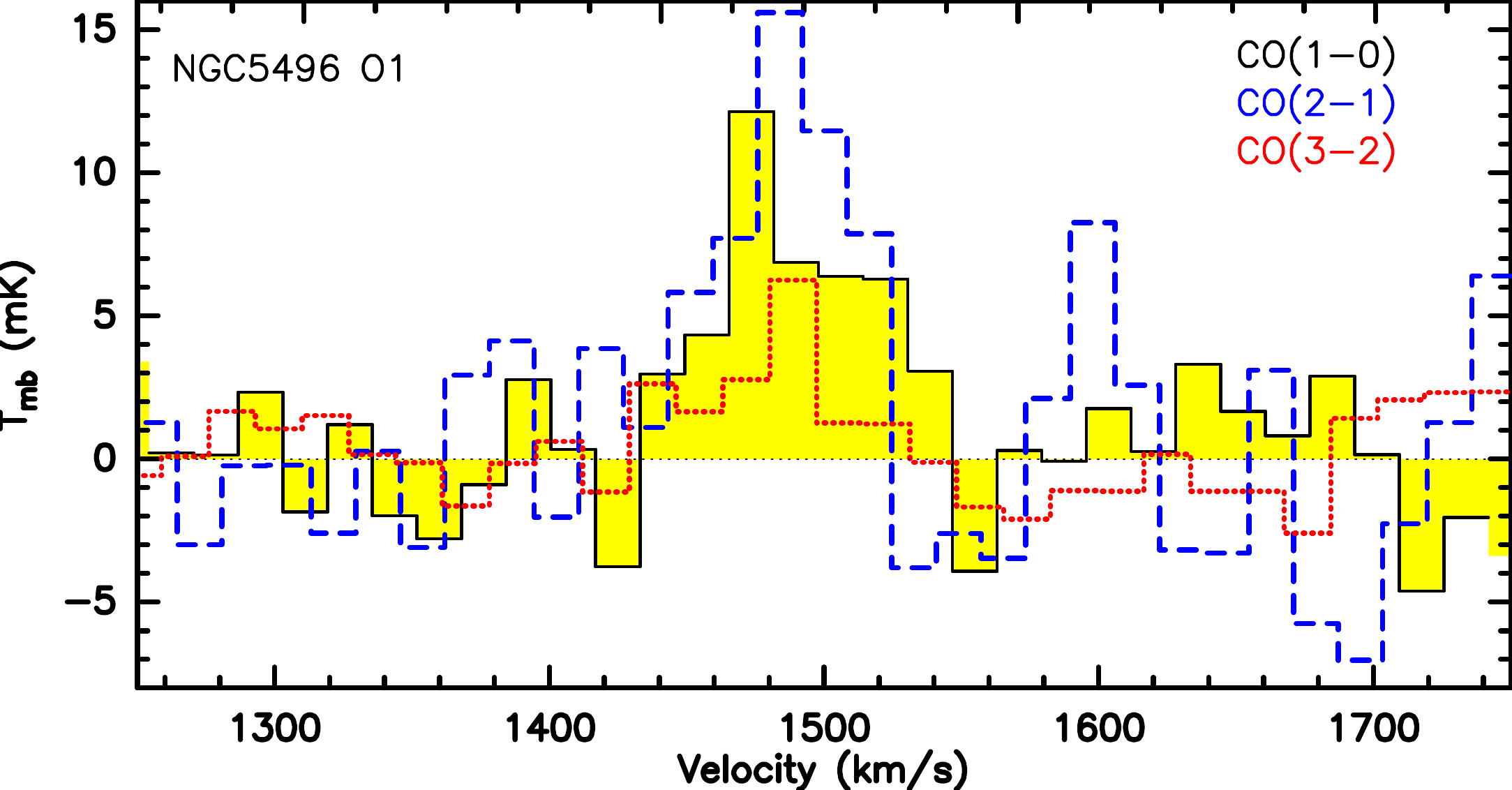}
 \includegraphics[width=\textwidth]{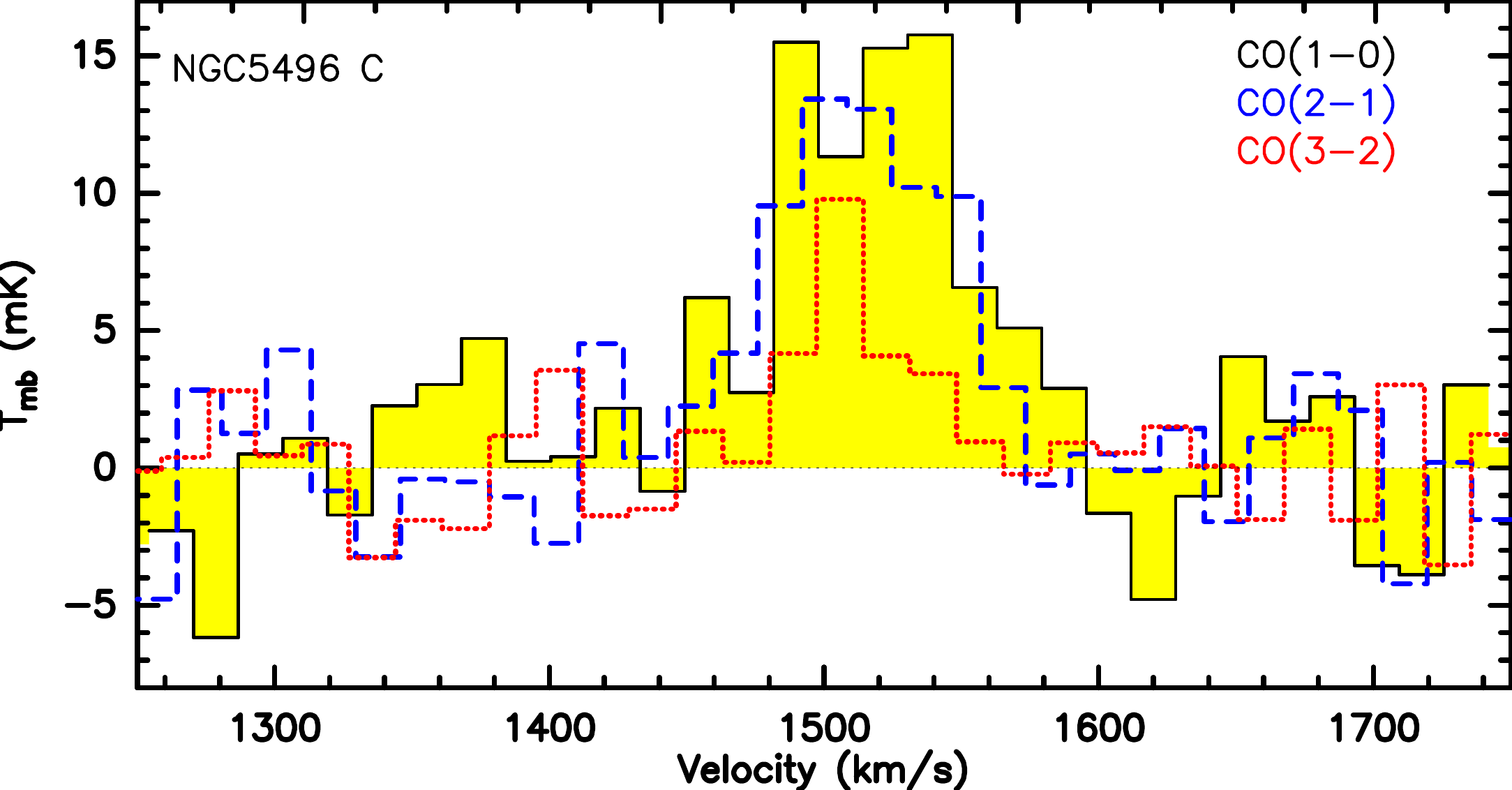}
 \end{minipage}
 \begin{minipage}[c]{0.32\textwidth}
 \includegraphics[width=\textwidth]{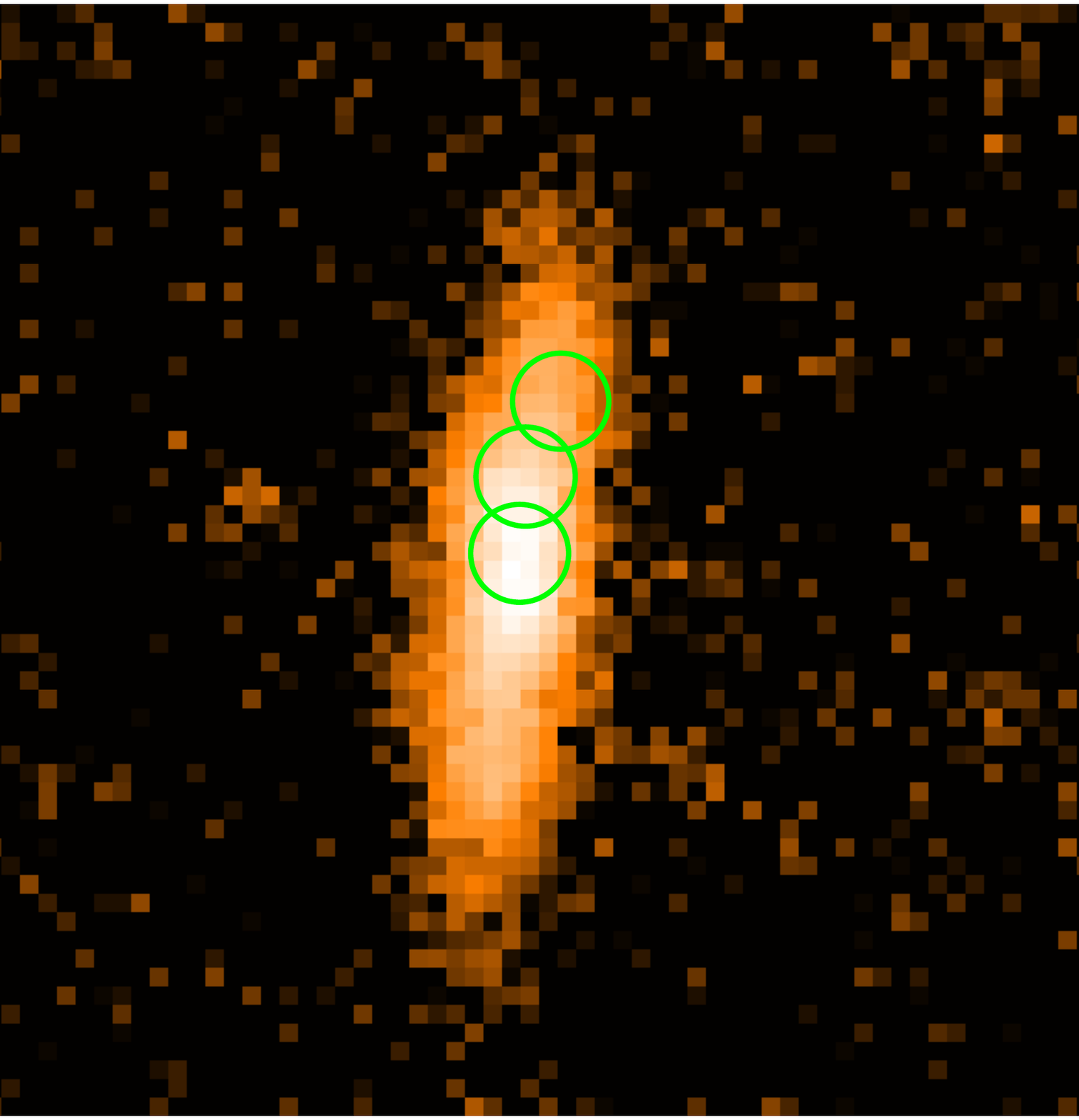}
 \end{minipage}
 \end{minipage}
 \caption{\label{spectraF}{\bf Left:} NUV/r/Z composite image of NGC5584 (top) and NGC5496 (bottom) with the CO pointings shown in yellow, representing the FWHM of the \COa\ beam (22\asec). {\bf Centre:} Spectra on the \tmb\ scale for each pointing, all three transitions are included in each panel. {\bf Right:} The 250\mic\ image (un-smoothed) with the CO pointings shown.}
\end{figure*}

\begin{figure*}
\begin{minipage}[c]{0.32\textwidth}
\includegraphics[width=\textwidth]{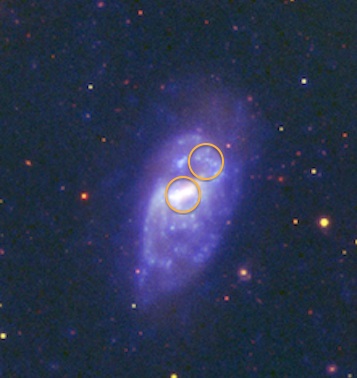}
\end{minipage}
\begin{minipage}[c]{0.32\textwidth}
\includegraphics[width=\textwidth]{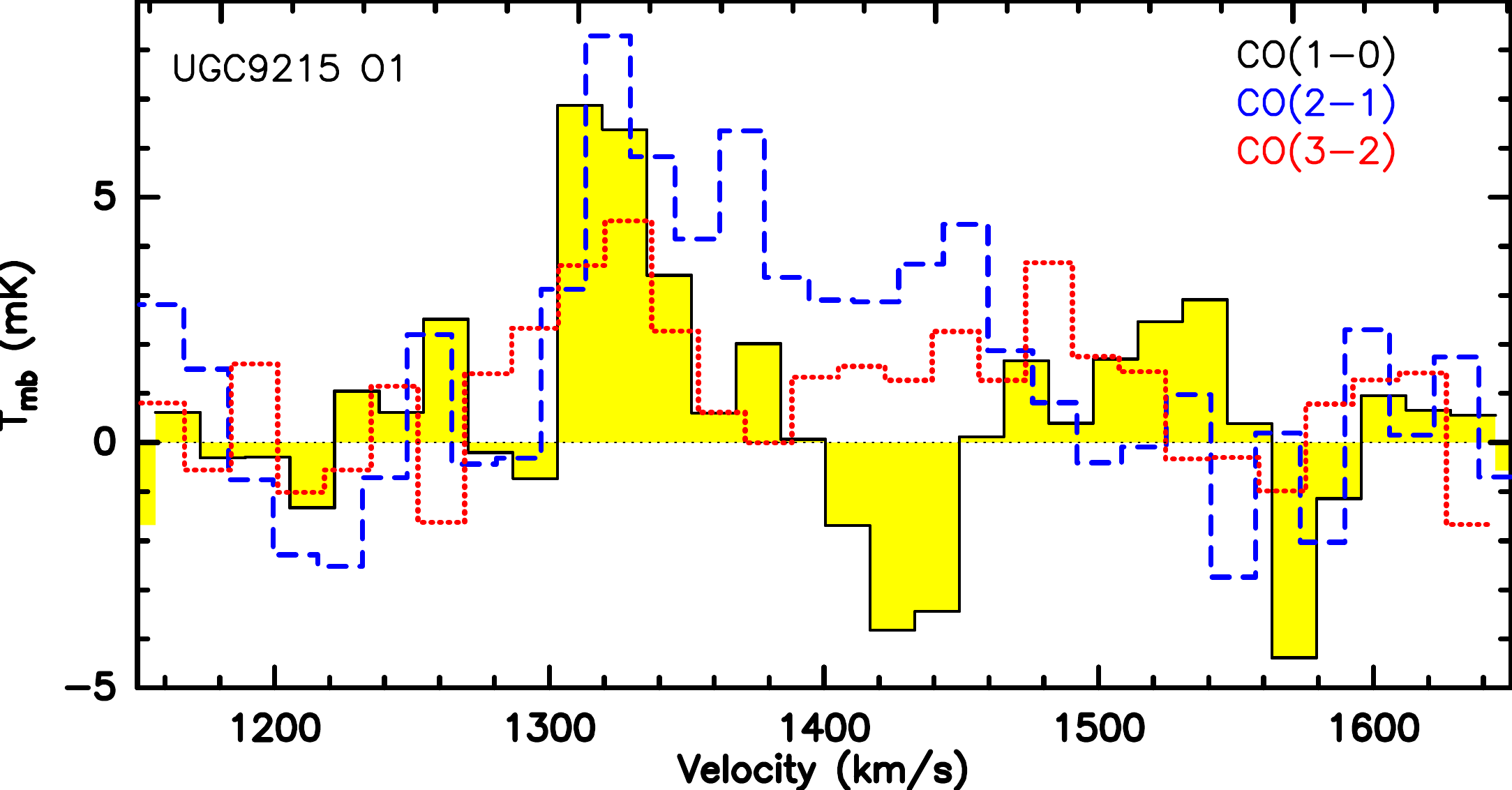}
\includegraphics[width=\textwidth]{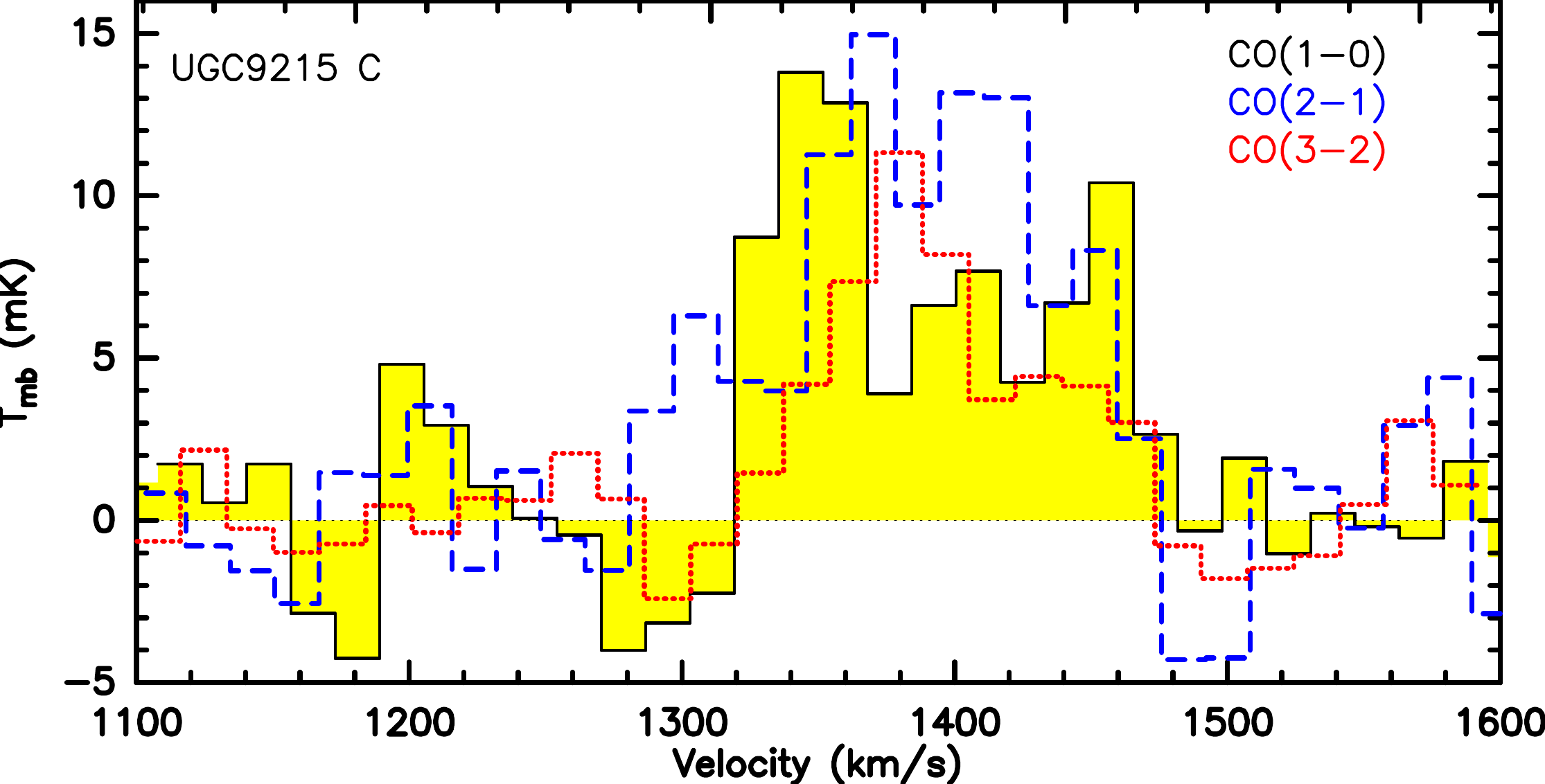}
\end{minipage}
\begin{minipage}[c]{0.32\textwidth}
\includegraphics[width=\textwidth]{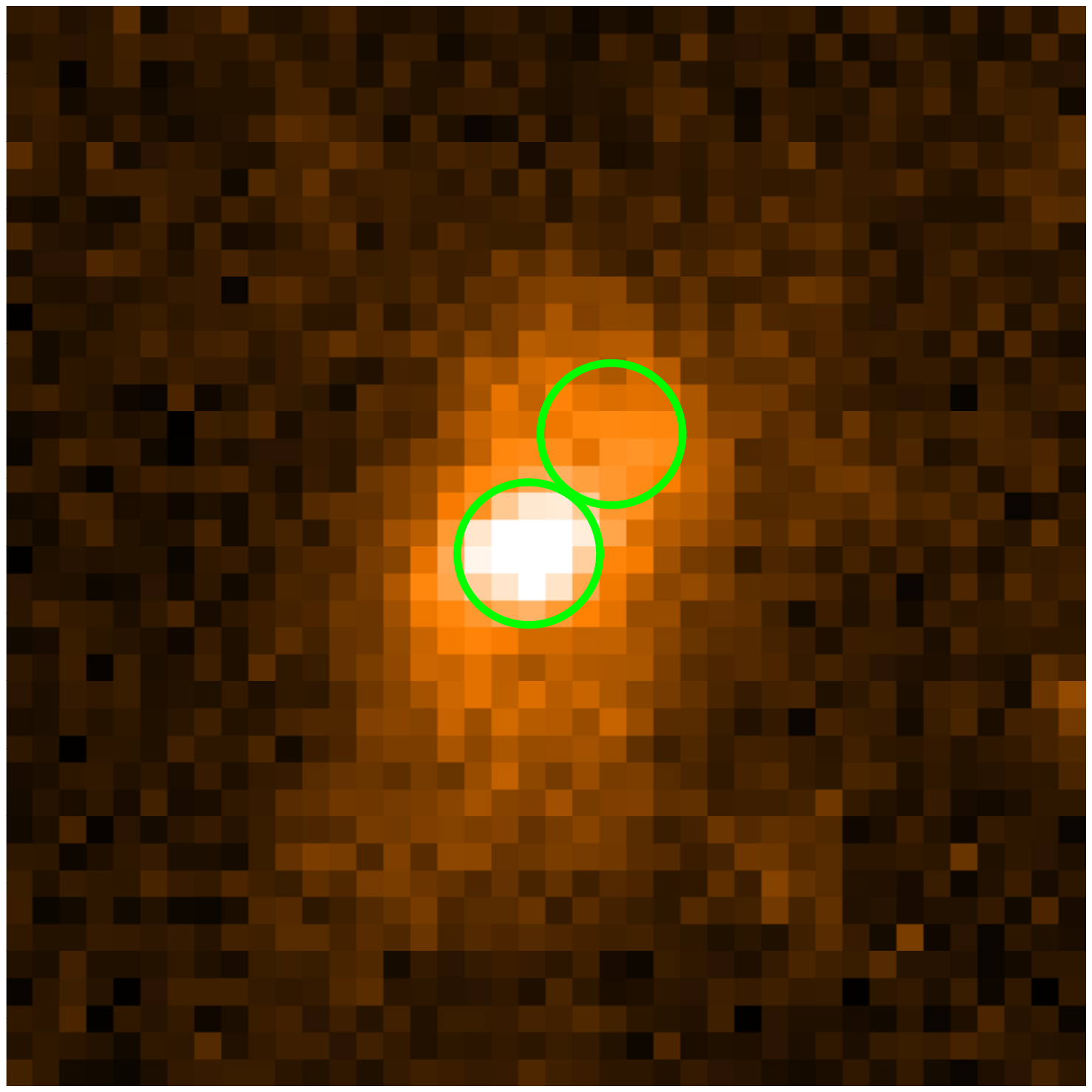}
\end{minipage}
\end{figure*}

\renewcommand{\thefigure}{\arabic{figure}}
\addtocounter{figure}{-1}

\begin{figure*}
\begin{minipage}[c]{0.32\textwidth}
\includegraphics[width=\textwidth]{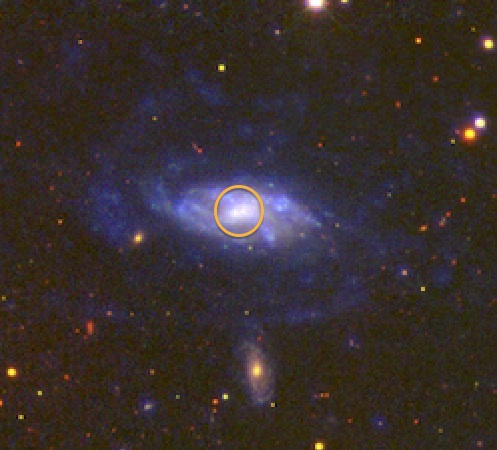}
\end{minipage}
\begin{minipage}[c]{0.32\textwidth}
\includegraphics[width=\textwidth]{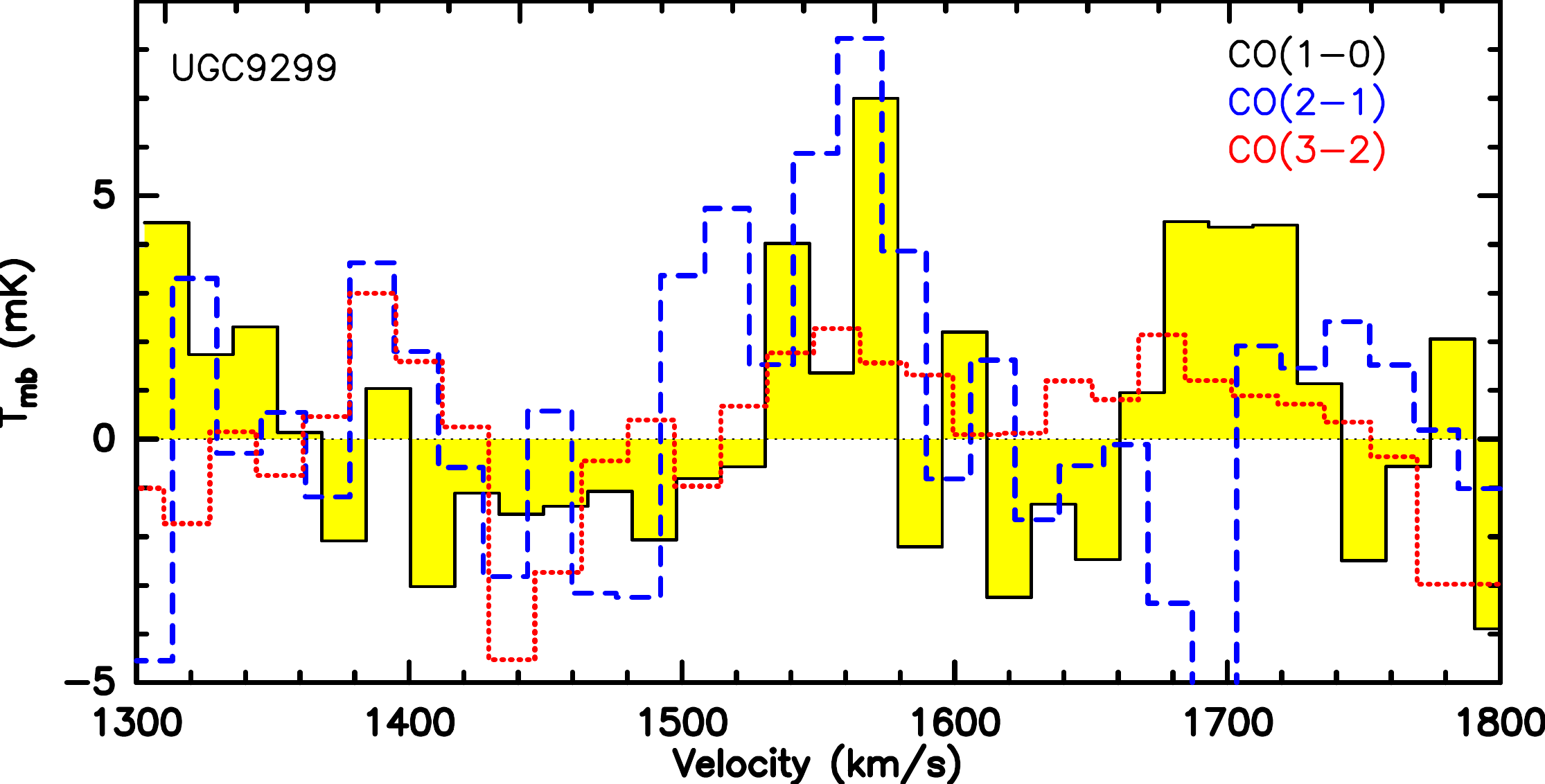}
\end{minipage}
\begin{minipage}[c]{0.32\textwidth}
\includegraphics[width=\textwidth]{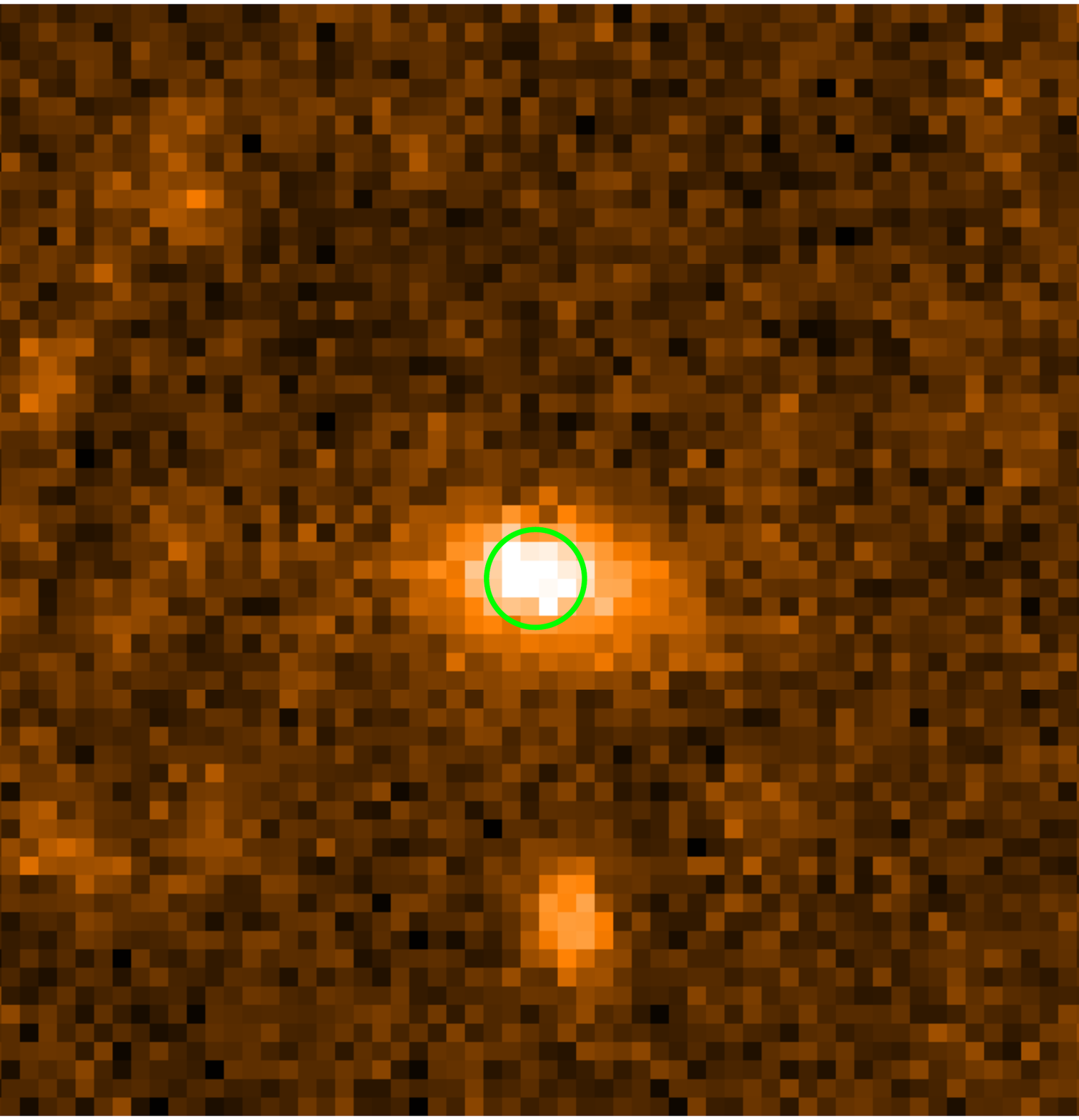}
\end{minipage}
\caption{\label{spectraF} (Cont.) {\bf Left:} NUV/r/Z composite image of UGC9215 (top) and UGC9299 (bottom) with the CO pointings shown in yellow, representing the FWHM of the \COa\ beam (22\asec). {\bf Centre:} Spectra on the \tmb\ scale for each pointing, all three transitions are included in each panel. {\bf Right:} The 250\mic\ image (un-smoothed) with the CO pointings shown.}
\end{figure*}

\renewcommand{\thefigure}{\arabic{figure}}

\section{CO and molecular gas properties}
\label{COpropsS}
The CO measurements in Tables~\ref{COdataT} and~\ref{COpropsT} are
used to provide diagnostics of the molecular gas in the 9 regions of
the four targets. The integrated line flux density gives a measure of the
molecular gas in the beam in \msun, as $\Mh= \aco \lcoa$, where \lcoa\ is given by:
\begin{equation}
\lcoa =  3.25\times10^7\,S_{10}\,\nu_{\rm{obs}}^{-2}\,D_L^2\,(1+z)^{-3}\,\,\rm{K\,kms^{-1}\,pc^{-2}}
\end{equation}
where $S_{10}$ is the integrated CO(1-0) flux density in Jy\kms,
$\nu_{\rm{obs}}$ is the observed frequency of the emission line in
GHz, $D_{\rm{L}}$ is the luminosity distance in Mpc and $z$ is the
redshift.
  
The conversion factor from CO luminosity to $H_2$ mass, \aco, is known
to be sensitive to metallicity
\citep{Bolatto2013,PPP2012,Wolfire2010,Glover2011,Leroy2011} although
it is not extremely sensitive for $Z > 0.5\,Z_{\odot}$ which applies
to all of our galaxies. To consider the impact of metallicity effects,
we make two estimates of \Mh\ for each region. For the first we use the
Milky Way value of $\aco = 4.3 \msun \,\rm{(K \kms pc^{-2})^{-1}}$
(including He) from \citet{Bolatto2013}, thus giving each region the
same CO-$\rm{H_2}$ relation. Secondly, we use the metallicity
dependent \aco\ of \citet{Wolfire2010}.\footnote{We also checked that
  using the \aco(Z) of \citet{Feldmann2012},\citet{Glover2011} or
  \citet{Genzel2015} did not change any of our conclusions: of all the
  relations, the \citet{Wolfire2010} \aco\ gave the maximum \Mh\ at a
  given metallicity.} This provides a unique scaling for each region
based on its local metallicity.

In general the CO emission is very weak with low peak temperatures
(5--30 mK (\tmb) and narrow line-widths\footnote{Two of these galaxies
  (NGC 5584 and NGC 5496) were also observed with the JCMT in the \COb\
  and \COc\ lines by \citet{Bourne2013}. The noise levels in the JMCT
  data were considerably higher but the results are consistent with
  our newer and deeper observations.}. In particular, the most massive
galaxy, NGC~5584, has very narrow lines in the outer regions of the
disk (FWHM $\sim$ 20-30\kms). Since the regions probed by the \COa\
beam are 2--3 kpc in diameter, such narrow lines indicate that the CO
is likely to be in small clouds with narrow intrinsic line-widths. The
implied molecular gas surface density (including He) averaged over
these 2--3 kpc regions ($\Sigma_{\rm H2,CO} = \aco\,I_{\rm{CO}}\,\cos
i$) is $\rm{\Sigma_{H2}}\sim 0.5-6\,\rm{M_{\odot}\,pc^{-2}}$. This is
$6-10\times$ lower than the density of {\em inter-arm\/} gas in M51
averaged over a similar area\citep{Colombo2014}. The values of
$\Sigma_{\rm H2,CO}$ for the individual pointings are listed in
Table~\ref{pressureT}.


\begin{landscape}
\begin{table}
\caption{\label{COdataT} Details of the CO measurements.}
\begin{tabular}{lccccccccccccc}
\hline
Pointing & R.A.       & Dec         & $v_{10}$ & $I_{10}$ & $\Delta v_{10}$ & $\sigma_{10}$ & $^{13}I_{10}$  & $I_{21}$ & $\Delta v_{21}$ & $\sigma_{21}$  & $I_{32}$ & $\Delta v_{32}$ & $\sigma_{32}$ \\
      & (J2000)    & (J2000)     & (\kms)     &  (K\kms)   & (\kms)     & (mK)           & (K\kms) &  (K\kms)   & (\kms)     & (mK)           &  (K\kms)   & (\kms)   & (mK)\\            
\hline
NGC5584 C  & 14:22:23.6 & $-$00:23:14 & 1636 & $2.11\pm0.15$ & 86 & 2.6  & $<0.16$ &$1.54\pm0.14$ & 71 & 2.6 & $0.81\pm0.11$ & 53 & 2.6\\        
NGC5584 O1 & 14:22:23.5 & $-$00:22:28 & 1573 & $1.07\pm 0.11$ & 35 & 4.8 & $<0.11$ & $1.04\pm0.14$ & 31 & 6.0 & $0.42\pm0.07$ & 28 & 3.5\\ 
NGC5584 O2 & 14:22:24.8 & $-$00:23:46 & 1710 & $1.48\pm 0.11$ & 32  & 4.5 & $<0.11$ & $1.39\pm0.15$ & 23 & 6.7 & $0.43\pm0.07$ & 23 & 4.0\\
\hline
NGC5496 C  & 14:11:37.8 & $-$01:09:24 & 1515 & $1.17\pm0.15$ & 73 & 3.1 & $0.28\pm0.06$ &$1.03\pm0.16$ & 71 & 3.0 & $0.45\pm0.09$ & 47 & 1.9\\
NGC5496 O1 & 14:11:37.7 & $-$01:09:03 & 1474 & $0.69\pm 0.09$ & 57 & 2.8 & $<0.14$ &$0.82\pm0.19$ & 49  & 4.4 & $0.32\pm0.07$ & 50 & 1.6 \\
NGC5496 O2 & 14:11:37.0 & $-$01:08:43 & 1456 & $0.63\pm 0.15$ & 100 & 1.8 & $<0.11$ & $0.63\pm0.12$ & 59 & 2.5 & $0.12\pm0.03$  & 23 & 1.3 \\
\hline
UGC9215 C & 14:23:27.3 & $+$01:43:34 & 1382 & $1.21\pm0.17$ & 87 & 3.2  & $<0.21$ &$1.35\pm0.12$ & 90 & 2.5 &  $0.78\pm0.09$ & 64 & 1.6 \\
UGC9215 O & 14:23:26.3 & $+$01:43:56 & 1319 & $0.33\pm0.07$ & 44 & 1.8 & $<0.09$ & $0.64\pm0.11$ & 75 & 2.0  & $0.23\pm 0.06$ & 51  & 1.5\\
\hline
UGC9299   & 14:29:34.6 & $-$00:01:06 & 1550 & $0.15\pm0.07$  & 22  & 2.4  & $<0.12$ &$0.47\pm0.09$  & 59  & 2.2 & $0.20\pm0.08$ & 75 & 1.5\\ 
           &  & &     &  ($<0.239$)$^a$ & (69)$^a$ &  &   &  ($0.31\pm0.10$)$^b$ & (36)$^b$ &    &               &        &    \\
\hline
\end{tabular}
{\flushleft \small{All intensity units are in the \tmb\ scale. $v_{10}$
    is the central velocity of the \COa\ line in the LSRK
    frame. $I_{CO}$ is the integrated intensity $\int{T_{mb}\delta v}$. $\Delta v_{CO}$ is the FWHM of the CO line as given by Eqn~\ref{FWHME}. $\sigma_{CO}$ is the spectral r.m.s. in the binned spectrum. $^a$ Measurement for UGC9299 made in 3 channels
    corresponding to the peak in \COb\ and \COc. Upper limit for
    UGC9299 from integrating across the FWZI of the main peak in the
    \COb\ line.  $^b$ \COb\ flux for UGC9299 measured in the same
    velocity range as \COa\ and \COc. Larger measurement not in
    parenthesis includes the blue-ward peak which is not evident in
    the other two lines.
}}\\
\end{table}
\end{landscape}

\begin{table*}
\caption{\label{COpropsT} Flux densities and line ratios of the molecular gas.}
\begin{tabular}{lccccccccccc}
\hline
Pointing     &  $S_{10}$ & $S_{21}$ & $S_{32}$ & $C_{21}$ & $C_{31}$ &  $R_{21}$ & $R_{31}$ & $R_{10}$  \\
           &  (Jy\kms) & (Jy\kms) & (Jy\kms) &    &   &   &       &        \\ 
\hline
N5584 C &  $10.53\pm0.75$       & $7.69\pm0.70$    & $24.9\pm3.4$ & 0.794 & 0.934  &  $0.58\pm0.07$ & $0.36\pm0.05$  & $>13.2$  \\
N5584 O1 &  $5.37\pm0.55$       & $5.20\pm0.70$      & $12.8\pm2.2$ & 0.676  & 0.877 & $0.66\pm0.11$ &$0.34\pm0.07$  & $>9.7$  \\
N5584 O2 &  $7.4\pm0.55$        & $6.94\pm0.75$     & $13.1\pm2.2$ & 0.847  & 0.943 &  $0.80\pm0.10$   &$0.27\pm0.05$  & $>13.5$  \\
\hline  
N5496 C  &  $5.87\pm0.75$      & $5.16\pm0.80$      & $13.8\pm2.8$ & 0.872  & 0.943 & $0.77\pm0.15$ & $0.36\pm0.09$   & $4.2\pm1.1$\\
N5496 O1 &  $3.47\pm0.45$     & $4.09\pm0.95$       & $9.8\pm2.2$ & 0.943  & 0.943 &   $1.12\pm0.30$   &$0.44\pm0.11$ & $>4.9$  \\
N5496 O2 &  $3.20\pm0.75$     & $3.17\pm0.60$       & $3.7\pm0.9$  & 1.00 & 1.00 & $1.00\pm0.30$ & $0.19\pm0.07$ & $>5.7$  \\
\hline
U9215 C  &  $6.05\pm0.85$    & $6.74\pm0.60$      & $24.1\pm2.8$ & 0.637  & 0.820 & $0.71\pm0.12$   & $0.53\pm0.10$  &   $>5.8$  \\  
U9215 O1 &  $1.65\pm0.35$    & $3.21\pm0.55$      & $6.82\pm1.9$ & 1.087  & 1.08 & $2.11\pm0.58$ & $0.75\pm0.25$ & $>3.7$  \\
\hline
U9299    &  $0.77\pm0.35$  & $1.54\pm0.50$  & $6.2\pm2.5$ & 0.719  & 0.877 & $1.49\pm0.84$  & $1.17\pm0.72$  & ..  \\
         &   ($<1.2$)      & $(2.33\pm0.75)$ &            &        &       &  ($>0.93$)     & ($>0.73$)      &     \\
\hline
\end{tabular}
\flushleft{\small{$S_{\nu}$/\tmb used are: 5.0 Jy/K for \COa, 5.01
    Jy/K for \COb\ and 30.85 Jy/K for \COc. $R_{21}$ and $R_{31}$ are
    the integrated line temperature ratios of the \COb/\COa\ and
    \COc/\COa\ lines respectively. $R_{21}$ and $R_{31}$ have been
    corrected for the mismatched beam sizes using the multiplicative
    factors $C_{21}$ and $C_{31}$ as described in the text. $R_{10}$
    is the ratio of $^{12}I_{10}/^{13}I_{10}$.}}
\end{table*}

\subsection{Line ratios}
The line luminosity or brightness temperature ratio provides some
diagnostic of the excitation of the molecular gas {\em averaged over
  the beam area\/}. For a point source sampled with the same beam size
in both transitions, or a very extended source, this is equivalent to
the ratio of line integrated intensities in \tmb\ units, such that
$\lcob/\lcoa = \tbb/\tba = I_{21}/I_{10}$. The 22\asec beam of IRAM at
115-GHz probes a physical scale of 2--3 kpc at the distance of these
sources, meaning that we cannot assume either a point source or a
uniformly extended source geometry. In order to interpret our line
ratios measured with different beams ($\COa= 21.5\asec, \COc=
17.5\asec, \COb= 10.7\asec$) we have to make corrections based on
assumptions about how the gas is distributed.


There are various ways in which corrections for mismatched beams are
attempted in the literature
\citep[e.g.][]{Braine1992,Grossi2016,Cormier2014,Hunt2017}. Following
\citet{Hunt2017}, we used the 160-\mic\ maps from our {\em Herschel}
PACS imaging to make a direct estimate of the relative surface
brightness of 160\mic\ emission within 22\asec, 18\asec and 11\asec
beam sized regions at the locations of our pointings. This assumes
that the CO distribution will follow that at 160\mic. Using this
method gives correction factors of 0.64--1.09\footnote{The correction
  can be greater than unity because a larger beam placed at the edge
  of an exponential distribution will see a higher surface brightness
  than a smaller beam, as the large beam samples closer to the peak of
  the exponential} for $R_{21}$ and 0.82--1.08 for $R_{31}$, which are
listed in Table~\ref{COpropsT}.\footnote{Instead, if we assume that
  the molecular gas is distributed as an exponential disk with
  scale-length $\sim 0.2\,R_{25}$ \citep{Young1995,Leroy2009,Kuno2007}
  we find slightly smaller but comparable values (0.58--1.03 for
  $R_{21}$).}

\begin{figure*}
\includegraphics[width=0.46\textwidth]{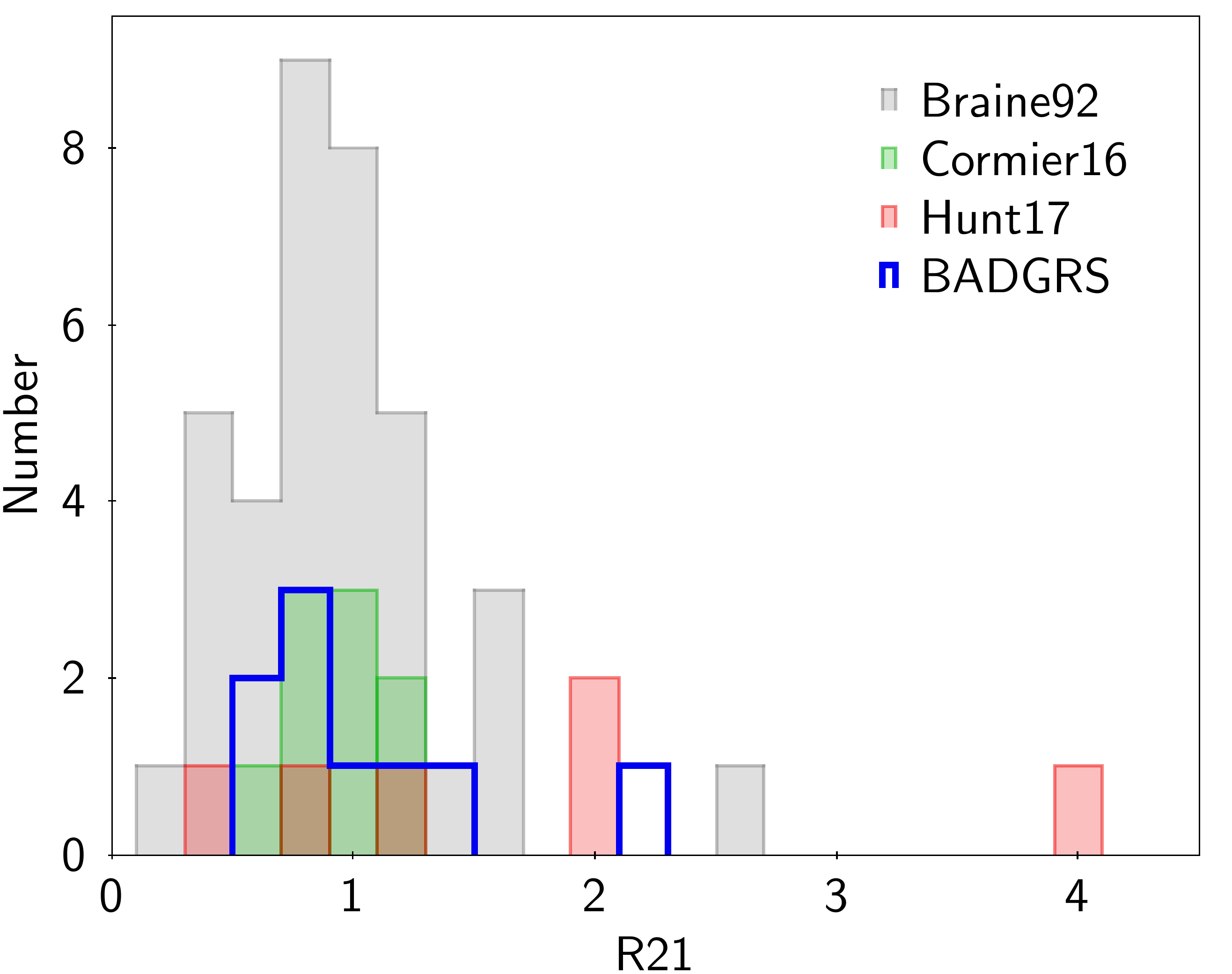}
\includegraphics[width=0.46\textwidth]{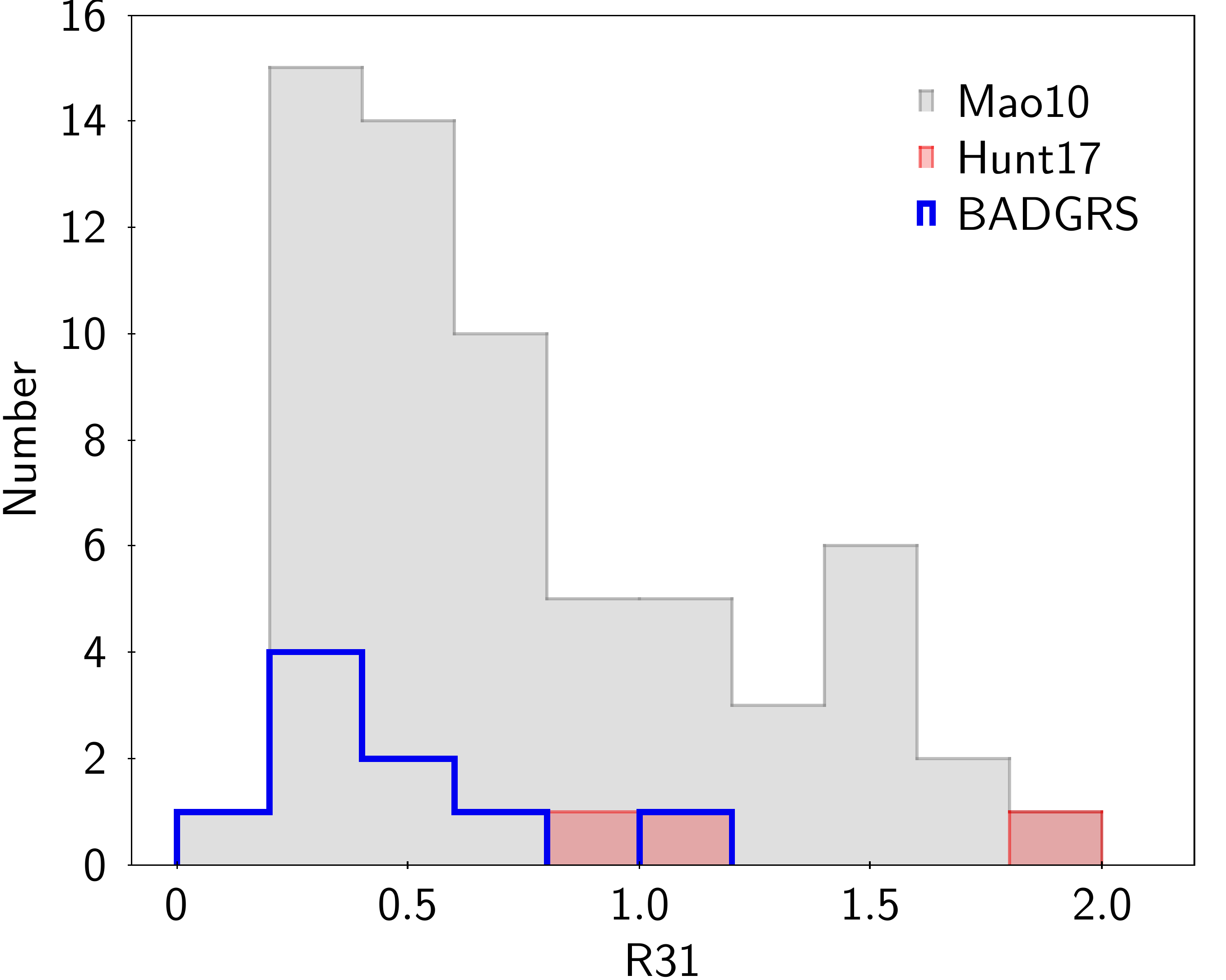}
\caption{\label{lineratiosF}{\bf (a)} Distribution of $R_{21}$ and {\bf (b)} $R_{31}$
  for BADGRS compared to literature samples of normal spirals from
  \citet{Braine1992} and \citet{Mao2010} and low metallicity galaxies
  from \citet{Hunt2017}. The BADGRS have similar $R_{21}$ ratios as
  other samples, but are less excited in the $R_{31}$ ratio compared
  to the low metallicity galaxies.}
\end{figure*}


The distributions of our corrected $R_{21}$ and $R_{31}$ values
compared to the literature are shown in Figure~\ref{lineratiosF}. The
$R_{21}$ ratios of BADGRS are consistent with other local samples of
spirals and dwarf galaxies, lying in the range 0.6--2.1. This
represents a variety of conditions: 
\begin{itemize}
\item{{\em Low excitation ($R_{21} =
  0.4-0.6$)}: found in quiescent inter-arm regions of external
galaxies and the outskirts of GMCs and dark clouds
\citep{Koda2012,Sakamoto1995,Sakamoto1997,Nishimura2015}. Low ratios
can be produced by low density gas or cold gas with $T_k <10$ K.}
\item{{\em
  Normal star forming ($R_{21} = 0.7-1.0$)}: found in the centres of
giant molecular clouds and typical star forming regions in nearby
galaxies
\citep{Braine1992,Sakamoto1994,Sakamoto1997,Koda2012,Leroy2009,Leroy2013,Cormier2014,Cormier2016}.} 
\item{{\em
  High excitation ($R_{21} \geq 1$)}: seen rarely in regions such as
the Galactic centre \citep{Sawada2001} and some star-bursting galaxies
\citep{Braine1992,Hunt2017}. To excite the gas to $R_{21}>1$ the
conditions could be warm, dense and optically thin with $\tau_{21}<5$
\citep{Sakamoto1997}. It is also possible to have such high ratios if
the gas is in optically thick dense clumps and heated externally by a
strong radiation field.}
\end{itemize}

The errors on the high $R_{21}>1$ ratios in NGC~5496~O1,O2 and UGC~9299 are large enough that they are
still consistent with a `normal' $R_{21}=0.7-1.0$ value, however
UGC~9215~O1 appears to be significantly excited with $R_{21}=2.11\pm0.58$. 

The $R_{31}$ ratios in Figure~\ref{lineratiosF}(b) are on average
lower than the samples of \citet{Mao2010} and \citet{Hunt2017}. The
lowest ratios ($0.2 < R_{31} < 0.3$) are in the outer disk regions
(O2) of the two most massive galaxies, NGC~5584 and NGC~5496. At face
value this low ratio indicates very cold ($\sim 10-20$ K) and/or low
density ($n \sim 1000-2000 \,\rm{cm^{-3}}$) clouds similar to those
found in the quiescent disk regions of external galaxies
\citep{Zhu2009,Wilson2009}, however, the $R_{21}$ values at these
locations are relatively high. The spectra in Figure~\ref{spectraF}
show that in these regions $\Delta v_{32} < \Delta v_{10}$, which
indicates that the warm, dense gas is located in more compact regions
than the cooler, more diffuse gas that dominates the J=1-0
emission. We believe that it is this different filling factor for the
two transitions which leads to a low beam averaged $R_{31}$. This is
especially evident in the NGC~5496~O2 region, where the \COc\ line is
very much narrower than the \COa\ line. This is a highly inclined
galaxy and the large sight-line through the disk exacerbates this
effect.
 
The regions with the highest $R_{21}$ and $R_{31}$ are those in the
outer disk of UGC~9215 and in UGC~9299. That the $R_{31}$ ratios are also
high (and less prone to uncertainty due to the closer match of the
beam areas), supports the earlier conclusion of more excited, dense
gas typical of star forming regions. The centre of UGC~9215 has an
intermediate value of $R_{31} \sim 0.53$ while for UGC~9215~O1 and
UGC~9299 the ratio is $R_{31} > 0.70$, which is more typical of
star-bursts, LIRGs and actively star forming regions where $R_{31} \sim
0.5-0.8$
\citep{Yao2003,Narayanan2005,Mao2010,PPP2012,Bauermeister2013}. We
find no trend for the $R_{31}$ ratio to be higher in the centre of
these galaxies compared to further out in the disk.

The isotopologue $\rm{^{12}CO/^{13}CO}$ ratio $R_{10}$ is useful for
breaking degeneracies in the low excitation CO Spectral Line Energy Distribution (SLED)
\citep[e.g.][]{PPP2012}. We present measurements or
upper limits for the ratio $R_{10}$ in Table~\ref{COpropsT}, using the
\COa\ line-widths given in Table~\ref{COdataT}. Low values of $R_{10} = 5-10$ indicate a cold,
quiescent phase with high $\rm{^{12}CO}$ optical depth whereas higher
values of $R_{10} \geq 15$ favour lower optical depth, vigorous star
formation with turbulent and diffuse clouds.  We detect $^{13}$CO(1-0)
only in the center of NGC~5496 where we see clear signs of a cold, high
optical depth and quiescent molecular ISM with $R_{10}=4.2\pm1.1$. In
NGC~5584, the lower limit of $R_{10}>13$ suggests that the gas in this
galaxy may have lower optical depth or a lower $^{13}$C abundance. For
the other positions there are no useful limits on $R_{10}$.

\section{Dust and molecular gas properties}
\label{DustS}

\begin{table*}
\caption{\label{DustT}Dust and molecular gas masses.}
\begin{tabular}{lccccccccc}
\hline
Pointing & $S_{250}$ (CO) & $S_{250}$ (tot) & $T_c$ & \Md (tot) & \Md (CO) & $\rm{12+log(O/H)}$ & $\alpha_{Z}$ & $\rm{M_{H2}}$ & \Mh /\Md \\
     &  (Jy)         &    (Jy)         &  (K)  & (Log \msun) & (Log \msun)  &     &   & (Log \msun) &    \\
\hline
N5584 C  & 0.402 & 5.880  & $13.8\pm1.2$ & $7.80\pm0.06$ & 6.63 &  8.69 (3) & 4.3 & 8.00 (8.00) & 23 (23) \\  
N5584 O1 & 0.445 &      &&                               & 6.68 &  8.52 (7) & 6.0 & 7.71 (7.85) & 11 (16)\\
N5584 O2 & 0.336 &      &&                               & 6.56 &  8.64 (4) & 4.3 & 7.85 (7.85) & 19 (19) \\     
\hline
N5496 C  & 0.500 &  2.513 &  $13.0\pm0.7$ & $7.53\pm0.05$ & 6.83 & 8.63 (1) & 4.3  & 7.67 (7.67)  &  7 (7)\\ 
N5496 O1 & 0.374 &        &&                              & 6.70 & 8.34 (4) & 10.3 & 7.44 (7.82) &  5 (13)\\
N5496 O2 & 0.165 &        &&                              & 6.35 & 8.32 (1) & 10.4 & 7.40 (7.78) &  11 (27) \\
\hline
U9215 C  & 0.492  & 2.000 & $13.8\pm1.4$ & $7.28\pm0.10$ & 6.67 & 8.44 (1) & 8.0  & 7.62 (7.89) &  9 (17) \\  
U9215 O1 & 0.219 &        &&                             & 6.32 &  8.36 (2)  & 10.1  & 7.06 (7.42) & 5 (13) \\
\hline
U9299    & 0.184  & 0.560 & $14.6\pm0.7$ & $6.74\pm0.05$  & 6.26 &  8.47  & 7.0 & $7.00$ ($7.21$) & $6$ ($9$) \\
\hline
\end{tabular}
\flushleft{\small{$S_{250}$(CO) is the 250-\mic\ flux in the area of
    the 115-GHz IRAM beam. $T_c$ is the cold dust temperature from the two component
    MBB fit. \Md(tot) is the total dust mass using the aperture
    250-\mic\ flux and the two component SED fit. \Md(CO) is the
    inferred dust mass using $S_{250}$(CO) and the global SED parameters. We
    use a dust mass opacity coefficient of $\kappa_{250}=0.56
    \,\rm{m^2\/kg^{-1}}$ from \citet{Planck2011xix}. Metallicities are
    taken from \citet{deVis2017CE} and \citet{Riess2011} and are
    quoted in the O3N2 calibration of \citet{Pettini2004}. The number
    of H{\sc ii} regions which have been averaged to get this
    metallicity are in parentheses. \aco\ is chosen to be either the MW
    value of 4.3 \msun $\rm{(K \kms pc^{-2})^{-1}}$ or a metallicity
    dependent value listed as $\alpha_Z$ taken from
    \citet{Wolfire2010}. All \aco\ include the contribution from
    He. \Mh\ values using $\alpha_Z$ are in parentheses following those
    derived from the MW \aco\ value.}}
\end{table*}

Dust emission usually shows a strong correlation with CO emission, and
by inference, $\rm{H_2}$ content
\citep{Young1995,Dunne2000,Dunne2001,Foyle2012,Corbelli2012,Scoville2014,Grossi2016,Hughes2017}.
This relationship has been exploited by many studies at both high and
low redshift which use dust as an alternative tracer for molecular
gas. The applications range from resolved studies in the Milky Way and
nearby galaxies, where dust emission is used in combination with H{\sc
  i} and CO to derive \aco\ factors or measure total $\rm{H_2}$
content
\citep{Dame2001,Draine2007,Bot2010,Roman-Duval2010,Gratier2010,Bolatto2011,Leroy2011,Sandstrom2013,SmithM2012,Shi2014}
to the potential use of dust emission at higher redshift as a
substitute for CO altogether
\citep{Magdis2012,Rowlands2014,Scoville2014,Genzel2015,Scoville2016,Hughes2017}. Given
the increasing use of dust as a tracer for gas, it is important to
understand if there are instances where that assumption breaks
down. In this section we will compare the dust mass with the molecular gas mass derived from the CO data.

\subsection{SED fitting and dust mass estimation}
The {\em Herschel}-ATLAS data are used to estimate the dust properties
of the four galaxies, using a two-temperature modified black body to
describe the SED. The total fluxes for NGC~5584 and NGC~5496 are taken
from \citet{deVis2017}, while for UGC~9215 and UGC~9299 updated FIR
photometry was measured from the maps using manually defined apertures
to avoid contamination from background sources. The {\em Herschel-}
SPIRE photometry includes the $\rm {K_{col}P}$ and $\rm{K_{beam}}$
corrections\footnote{See section 5 of
  \url{http://herschel.esac.esa.int/Docs/SPIRE/spire_handbook.pdf}}
(1.019, 1.0019, 1.005 at 250, 350 and 500-\mic) and aperture
corrections as described in \citet{Valiante2016}. For {\em Herschel}
PACS we did not apply a colour correction as the SED shapes were close
to $F_{\nu}\propto \nu^{-1}$ for the 100 and 160\mic\ points. We also
used the {\em IRAS} 60-\mic\ flux in the SED fit and constrained the
maximum warm dust contribution by using the {\em WISE} 22-\mic\ flux as
an upper limit. More details of the photometry are in
\citet{Clark2015} and \citet{deVis2017}. We use a value for the dust
mass absorption coefficient of $\kappa_{250} = 0.56
\,\rm{m^{2}\,kg^{-1}}$, which gives masses a factor 1.59 times higher than
\citet{Clark2015} and \citet{deVis2017} (and for reference 1.4 times
      {\em lower\/} than other commonly adopted opacities from the
      \citealt{Draine2007} model). This updated coefficient is derived
      from the {\em Planck} 857-GHz measurement of the dust opacity
      per H nucleon assuming a dust-to-gas ratio of 1\%, and scaling
      to 250\mic\ using $\beta=1.75$ as measured for this part of the
      spectrum in \citet{Planck2011xix}. Of course, wherever we
      compare dust masses to other samples from the literature we
      scale all masses to the value of $\kappa_{250}$ used here.

\begin{figure}
\includegraphics[width=0.5\textwidth]{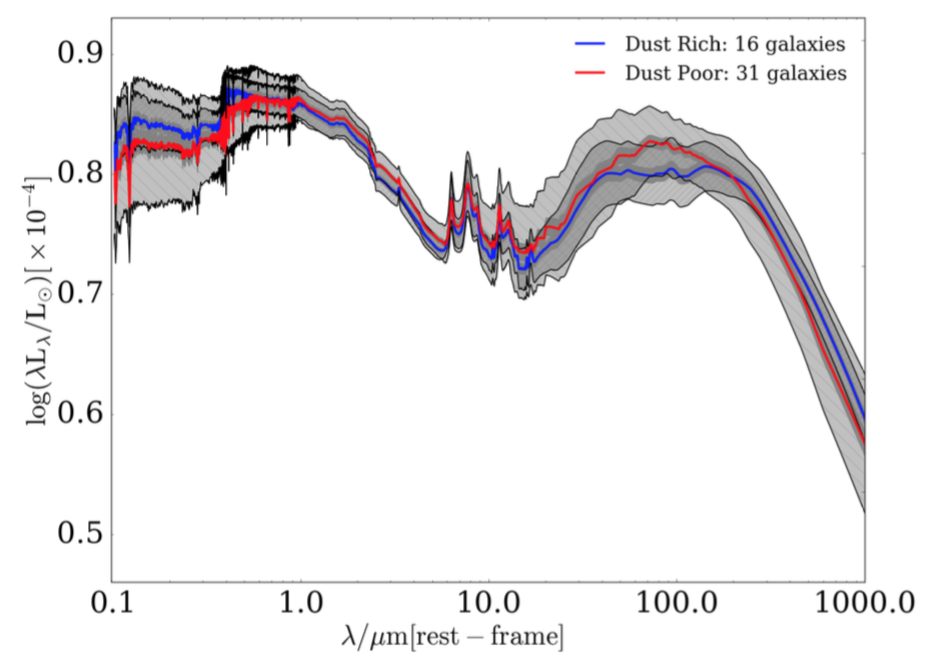}
\caption{\label{SEDstackF} Stacked best-fit SEDs from {\sc magphys}
  for BADGRS (blue) and non-BADGRS from the sample of
  \citet{Clark2015}. The BADGRS have broad and flat IR SEDs showing
  significant emission from cold dust. This figure is reproduced from \citet{Schofield2017}.}
\end{figure}

When well sampled, the dust SED between 60--1000\mic\ in star forming
galaxies is not well represented by a single temperature modified
blackbody because the dust is at a range of temperatures
\citep[e.g.][]{Devereux1990,Dunne2000,Dunne2001,Dale2002,Draine2007,Galliano2011,Remy-Ruyer2015,Hunt2015,Clark2015}. Warm
($>30$ K) dust is located near to regions of star formation while dust
in the diffuse ISM is heated by the interstellar radiation field
(ISRF) to cooler temperatures of 10--25 K.\footnote{Cold dust may also
  be present in dense starless cores, but this is not thought to
  dominate the mass of dust, at least in normal galaxies
  \citep{Draine2007}.} Whilst in reality the dust is at a range of
temperatures along the line of sight, a reasonable estimate of the
dust mass is obtainable by fitting a two temperature component
modified blackbody (2MBB), as described in \citet{Dunne2001}. The bulk
of the dust mass is contained in the diffuse, colder component and so
as long as the SED fitting method allows this to be separated from the
hotter components, we do not miss a large fraction of the dust
mass. This method gives similar dust masses to that of
\citet{Draine2007} who allow the dust to be heated by a radiation
field with a power law distribution. The errors on the temperatures
and dust masses are calculated using the $\chi^2$ distribution for the
SED fits for each parameter, and represent the 68\% confidence
interval. Our two component modified blackbody fits to the
60--500\mic\ photometry produces parameters very similar to those in
\citet{Clark2015}, and we repeat their finding that the diffuse dust
temperatures for the BADGRS sample are significantly cooler ($\sim
13-15$ K) than those in more massive spirals and low metallicity dwarf
galaxies (18-31 K) \citep{Remy-Ruyer2015,Clark2015,Hunt2015}. This is
illustrated in Fig.~\ref{SEDstackF} where the stacked best fit SEDs
from {\sc magphys} for BADGRS and non-BADGRS from the samples of
\citet{Clark2015,deVis2017} are compared. The BADGRS dust SED has a
much flatter peak and relatively more emission in the sub-millimeter
part of the spectrum where the cold dust dominates. A broader FIR SED
shape for lower stellar mass, lower metallicity galaxies has been
noted before and is thought to be due to a larger range in ISRF
intensities across the low attenuation galaxies, allowing dust to be
heated to higher temperatures
\citep{Ciesla2014,Cortese2014,Remy-Ruyer2015}. However, the ubiquitous
very cold diffuse dust temperatures we see in BADGRS have not been
found in other samples.

To estimate the dust mass within the same region as the CO emission we
use the 250-\mic\ map as it has the closest intrinsic resolution to the
CO data without sampling a larger beam area, and is also the highest
sensitivity map from {\em Herschel}. We convolve the 250-\mic\ map to
the same resolution as the \COa\ data\footnote{This is a small change
  as the 250-\mic\ beam is 18.5\asec and the \COa\ beam is 21.5\asec.}
and use the pixel value at the location of the CO pointing to give
$\rm{S_{250}(CO)}$.\footnote{The peak pixel flux is corrected by the
  factor 1/0.95 to account for the effects of pixelisation as
  described in the SPIRE manual.}  The dust mass in the area covered
by the IRAM beam at 115 GHz, $\rm{M_d(CO)}$, is then derived from

\begin{equation*}
\rm{M_d(CO) = (S_{250}(CO)/S_{250}(tot)) \times M_d(tot)}
\end{equation*}

This scaling simply assumes that the cool dust SED is similar on these
scales within the galaxy. These galaxies do not have large bulges and
therefore there is no expectation of a strong radial temperature
gradient for the diffuse dust component which dominates the dust
mass. We cannot perform the 2MBB fitting to the individual regions, as
these galaxies are unresolved by {\em IRAS} at 60-\mic.

The local dust masses in each CO region are listed in
Table~\ref{DustT} and are also compared to the local \Mh\ masses
derived from the \COa\ data. 

\subsection{Molecular gas-to-dust ratios and star formation efficiency}

\begin{figure*}
\includegraphics[width=0.45\textwidth,trim=0.8cm  0.7cm 1.1cm 0.7cm, clip=true]{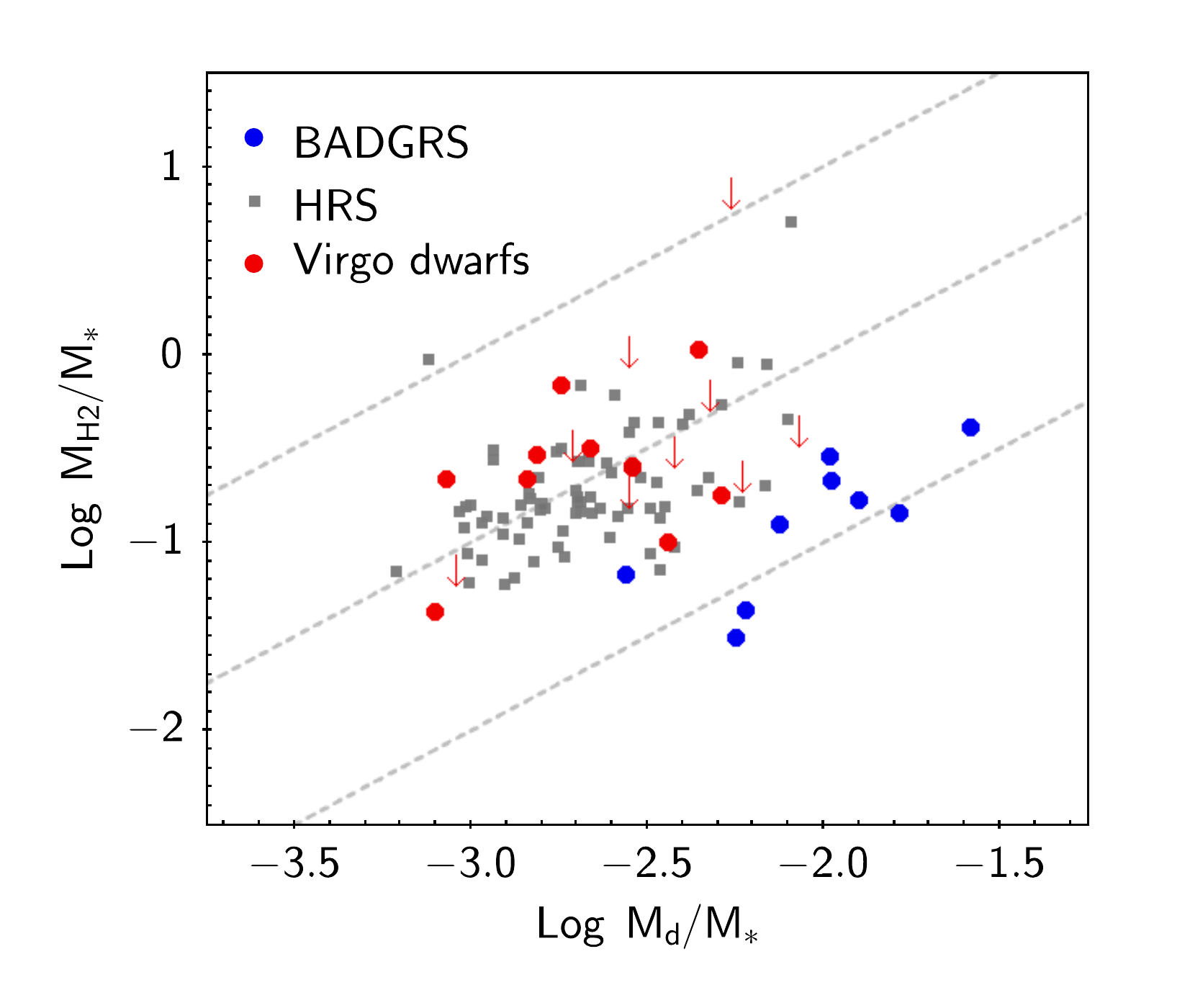}
\includegraphics[width=0.45\textwidth,trim=0.8cm  0.7cm 0.8cm 0.7cm, clip=true]{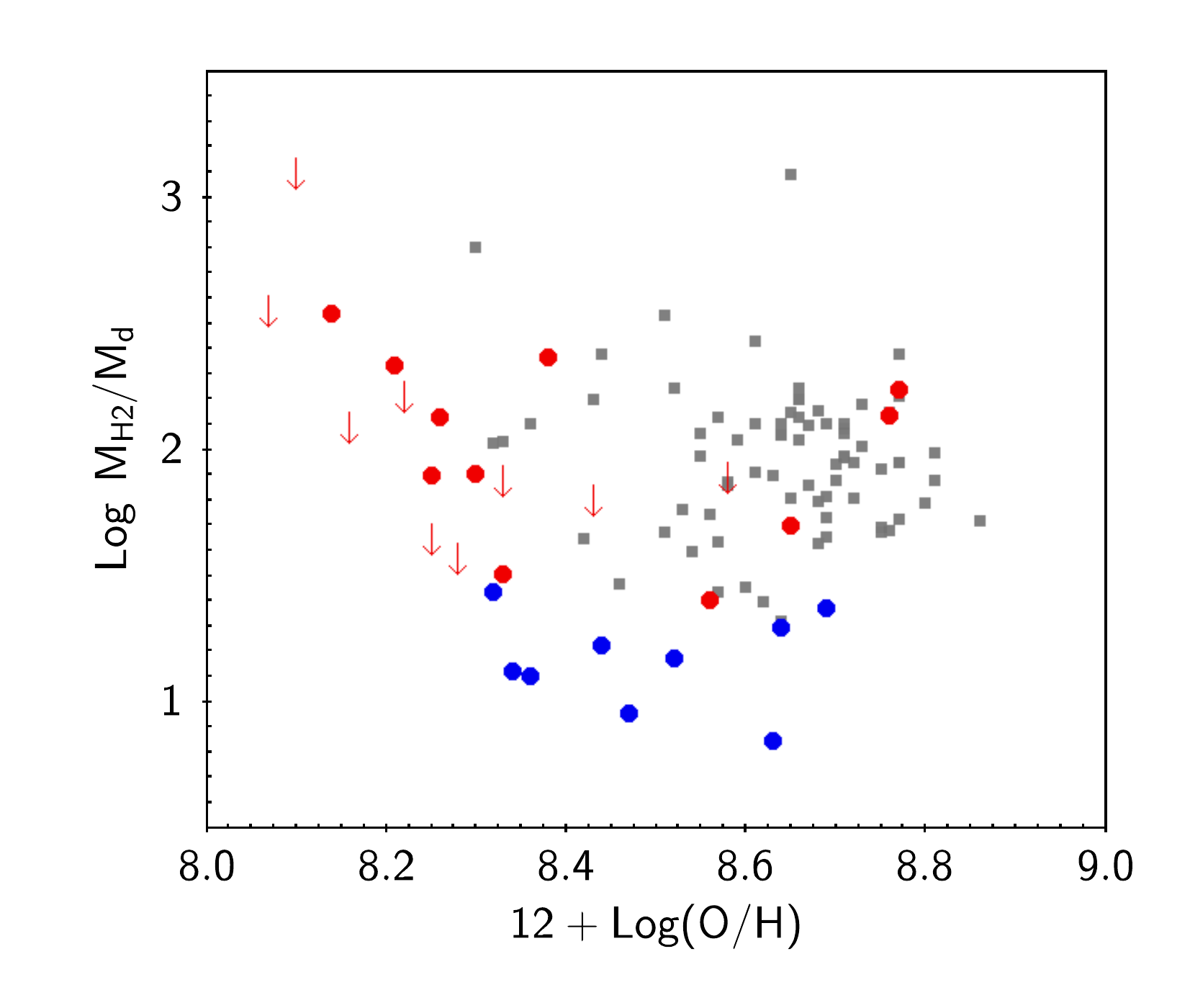}
\caption{\label{GrossiF} {\bf (a)} The correlation between dust and
  molecular gas mass having normalised by the common scaling factor of
  stellar mass. The dashed lines indicate constant \Mh/\Md\ ratios of
  10, 100 and 1000 (bottom to top). BADGRS are the most dust rich
  (relative to \Ms) galaxies and lie below the extrapolated trend for
  \Mh/\Md\ from the other samples. {\bf (b)} Molecular gas to dust
  ratio as a function of metallicity. There is no correlation, but
  BADGRS lie clearly below the average \Mh/\Md\ ratios for the other
  samples. Metallicities are all in the O3N2 calibration of
  \citet{Pettini2004}. Dust masses for the HRS are from {\sc magphys}
  \citep{deVis2017}, while dust masses for Virgo dwarfs are scaled
  from the $\beta=2$ version in \citet{Grossi2016} to be comparable to
  {\sc magphys} as follows. The HRS {\sc magphys} dust masses were
  compared to those determined by \citet{Grossi2016} using their
  single MBB $\beta=2$ method. A tight correlation with linear slope,
  and an offset of $\rm{Log\,(M_{d,MP}/M_{d,G16}) = 0.23}$ dex was
  found (once all were scaled to the same $\kappa_d$). This difference
  is due to {\sc magphys} fitting a more realistic temperature
  distribution which captures better the colder dust component. We
  scale the Virgo dwarf dust masses up by this amount to be consistent
  with HRS and BADGRS. All dust masses are scaled to a common
  $\kappa_{250}=0.56\, \rm{m^2\,kg^{-1}}$. }
\end{figure*}

For a metallcitiy dependent \aco, the \Mh/\Md\ ratios for BADGRS range
from 7--27 with a mean $\Mh/\Md=16$ (the mean using the $\alpha_{\rm{MW}}$
is 11). This is lower by a factor $\sim 10$ compared to the $\Mh/\Md
\sim 100-150$ seen in local spirals
\citep{Dunne2001,Leroy2011,Sandstrom2013}, as illustrated in
Figure~\ref{GrossiF} which compares the dust and molecular gas
properties for the BADGRS, HRS \citep{Boselli2014} and star forming
Virgo dwarfs \citep{Grossi2016}. The dashed lines in
Fig~\ref{GrossiF}(a) represent constant $\Mh/\Md$ ratios of 1000, 100
and 10 (top to bottom). Fig~\ref{GrossiF}(b) shows that BADGRS display
a much lower $\rm{H_2}$-per-unit-dust-mass at the same metallicities
as other galaxies. There is no obvious trend of \Mh/\Md\ with
metallicity.

\begin{figure}
\includegraphics[width=0.47\textwidth]{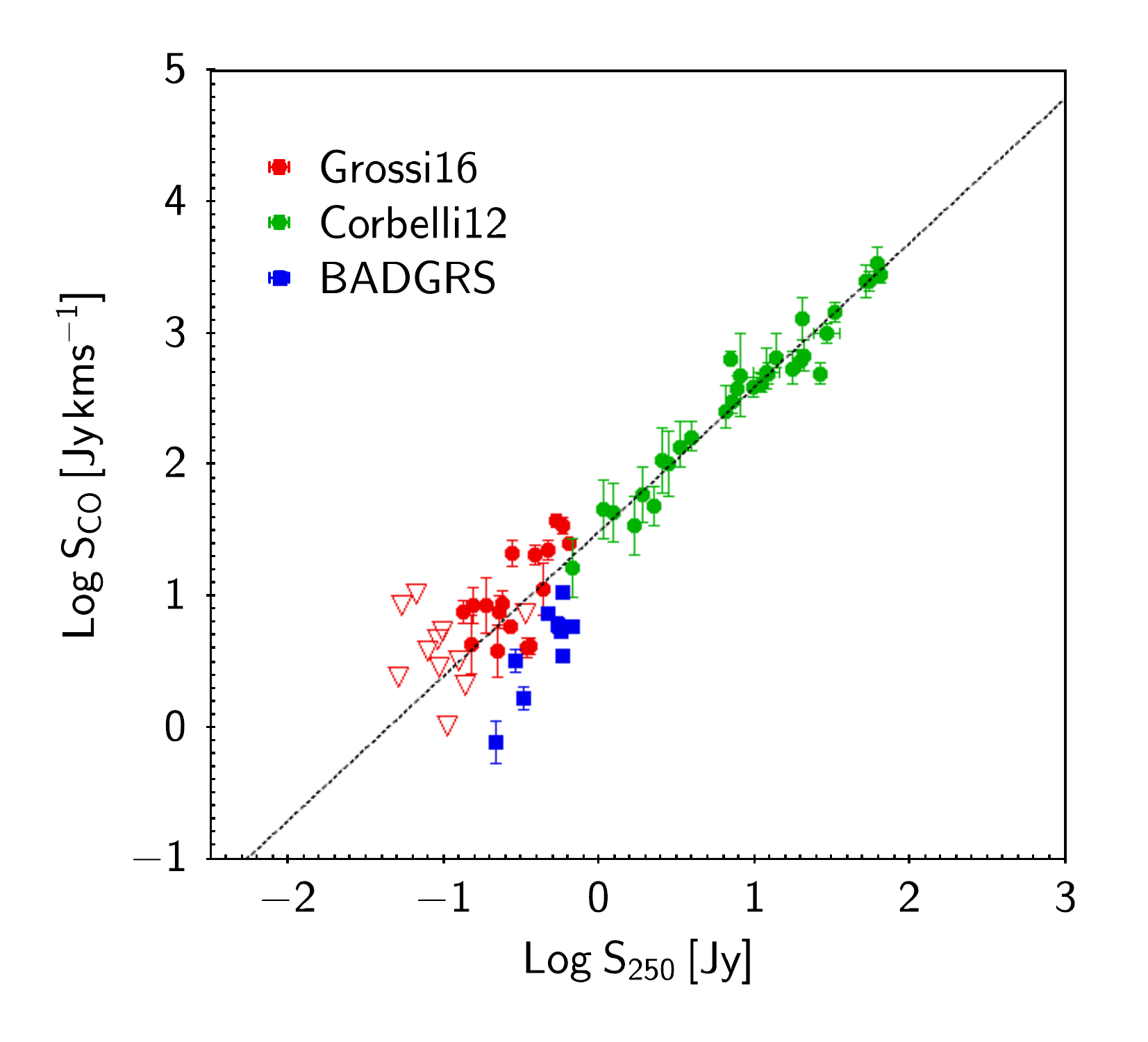}
\caption{\label{COS250F} The tight correlation of CO and sub-mm fluxes
  first presented in \citet{Corbelli2012} and updated in
  \citet{Grossi2016}. We reproduce the literature data here and
  compare to our BADGRS sample. The open triangles represent upper
  limits. The dotted line is the best fit from \citet{Grossi2016} to
  the red and green points. These fluxes are single IRAM CO fluxes in
  a 22\asec beam compared to the 250-\mic\ flux in an aperture with
  radius 18.5\asec as described in \citet{Grossi2016}. It can be seen
  that the BADGRS lie below the relationship found in other samples.}
\end{figure}

To be certain that these low \Mh/\Md\ ratios are not produced
purely by the conversions from observed quantities to dust and
$\rm{H_2}$ mass, we compare the observed quantities in
Figure~\ref{COS250F}. There is a tight relationship between CO
emission and 250-\mic\ flux in local galaxies from the HRS
\citep{Corbelli2012} and star forming dwarf galaxies in Virgo
\citep{Grossi2016}. Both the massive metal rich sources from the HRS
and the more metal poor Virgo dwarfs lie on the same
$F_{\rm{CO}}-F_{250}$ relation.\footnote{We followed the same
  prescription as \citet{Grossi2016} to estimate $S_{250}(beam)$ which
  is to use an aperture of radius 18.5\asec. Note this is not the same
  as our own values given by the convolution of the {\em Herschel}
  maps to the same resolution as the IRAM data, the \citet{Grossi2016}
  method results in higher dust fluxes.} The BADGRS are deficient in
CO emission at a given 250-\mic\ flux, meaning their anomalously low
gas-to-dust ratios are not artifacts of the methods used to convert from flux to mass.

\begin{figure*}
\includegraphics[width=0.46\textwidth, trim=0.0cm 0.0cm 0.0cm 0.0cm,
  clip=true]{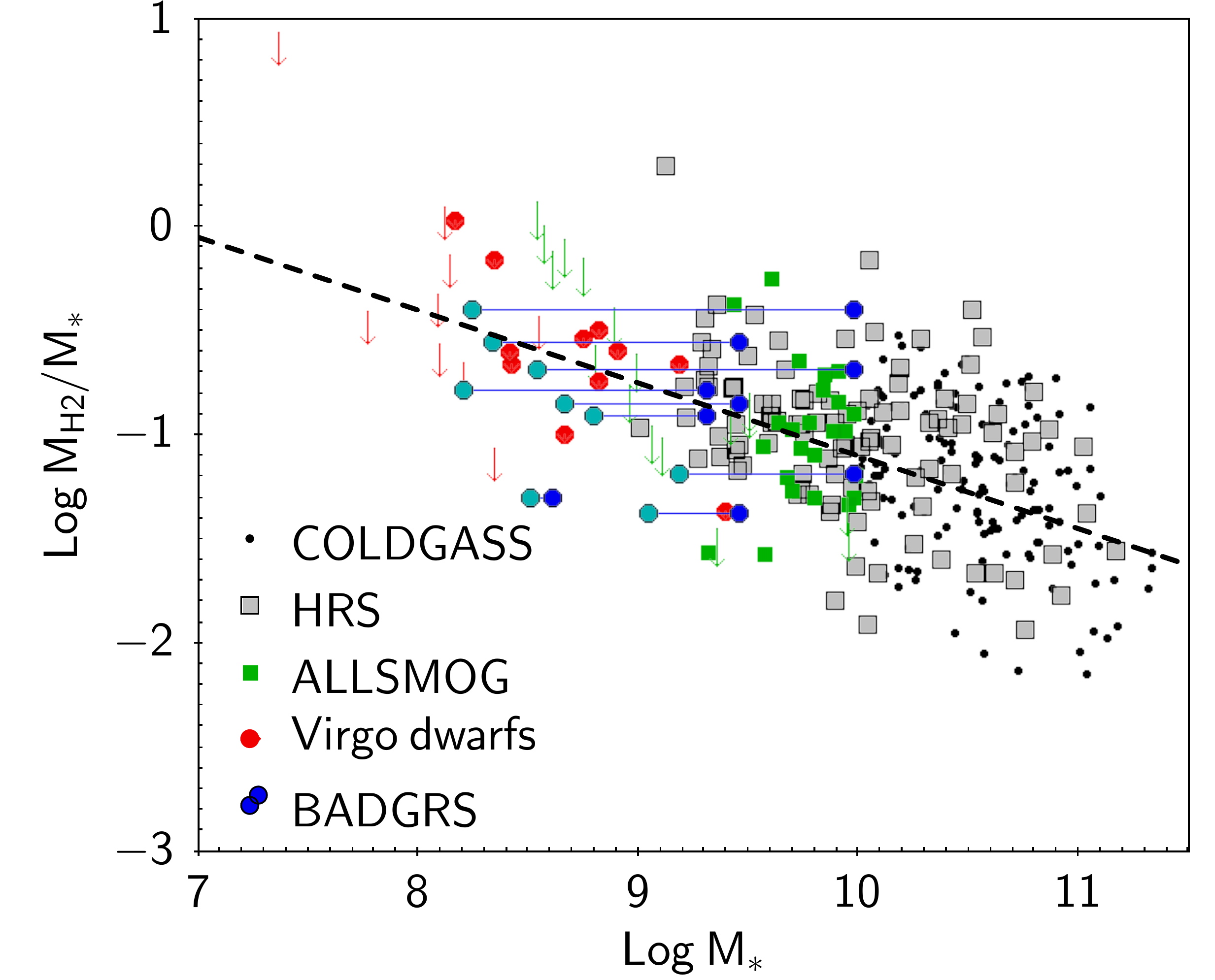}
\includegraphics[width=0.45\textwidth, trim=0.0cm 0.0cm 0.0cm 0.0cm,
  clip=true]{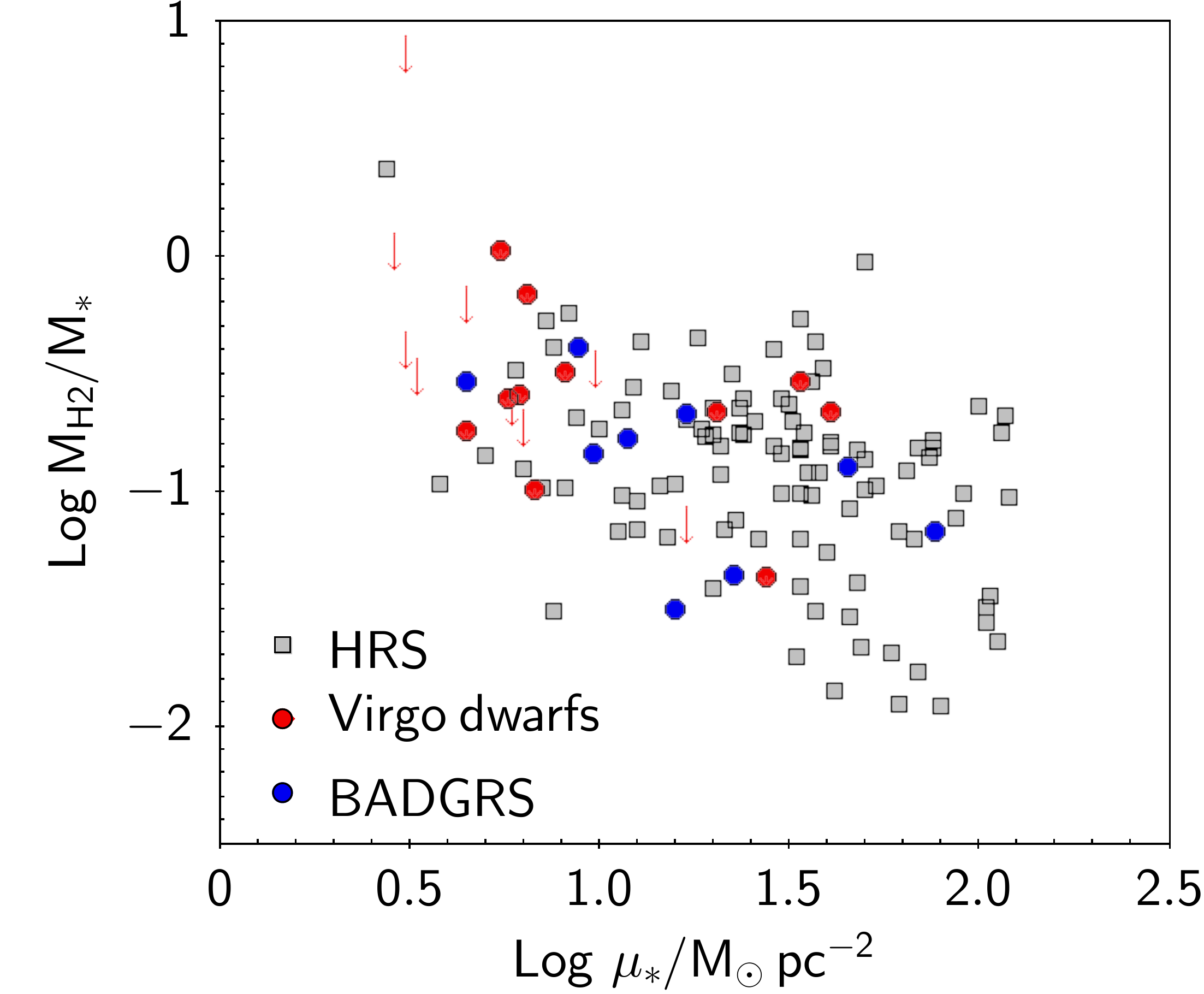}
\caption{\label{msmhF}a) \Ms--\Mh\ scaling relation with the
  relationship from \citet{Bothwell2014} for the ALLSMOG galaxies
  shown as black dashed line. All \Mh\ include He. BADGRS are shown
  both at their global stellar mass (dark blue circles) connected to
  the stellar mass in each pointing (cyan circles). When only the
  local values within the pointing are used (cyan) the BADGRS show a
  tight relationship but offset to lower \Mh/\Ms\ at similar stellar
  mass. Using the total stellar mass on the x-axis results in the dark
  blue points, which are comparable to the other samples of spirals
  and dwarfs, which are mainly selected based on their stellar
  masses. Only late type HRS galaxies with a beam filling fraction
  greater than 10 percent of the optical area are shown
  \citep{Boselli2014}, as the uncertainty of the extrapolation to
  total CO flux becomes too uncertain. The ALLSMOG points use a value
  of $r_{21}=0.7$ rather than $r_{21}=1$ as in \citet{Bothwell2014};
  this is more typical of the ratio observed in similar samples. The
  conversion from $\rm{L_{CO}-\Mh}$ for the BADGRS, Virgo dwarfs, HRS
  and ALLSMOG use the \citep{Wolfire2010} metallicity dependent
  \aco. The COLDGASS data points are from \citet{Huang2014} who use a
  matched 22\asec region for stellar mass and \Mh, and a constant
  \aco(MW) factor (they only include high stellar mass, high
  metallicity galaxies). Stellar masses for all the samples are
  estimated in ways which do not produce any large systematic
  offsets. b) \Mh/\Ms\ versus stellar mass surface density for those
  samples with available data. In this plot there are no differences
  between the BADGRS and other samples.}
\end{figure*}

In Figure~\ref{msmhF}(a) we show the scaling relation for molecular
gas and stellar mass. The \Mh/\Ms\ ratios for the BADGRS are for the individual
pointings; the stellar mass in the area of the IRAM beam is derived
from smoothed stellar mass maps created from the $g$ and $i$ SDSS
images and the \citet{Zibetti2009} stellar mass prescription. From the
literature we include the stellar mass selected COLDGASS
\citep{Saintonge2011,Huang2014}, ALLSMOG
\citep{Bothwell2014,Bothwell2016} and HRS \citep{Boselli2014} samples
and also the star forming Virgo dwarfs \citep{Grossi2016}. Adjustments
have been made to literature values in order to bring all measurements
onto common scalings, see caption to Fig~\ref{msmhF} for details. We
show two locations for the BADGRS: dark blue circles are using the
total stellar mass on the {\em x}-axis, cyan circles are using the
stellar mass within the IRAM beam. A better comparison, given our
resolved observations, is to use the stellar mass surface density on
the abscissa rather than total stellar mass. Figure~\ref{msmhF}(b)
shows more correctly that there is no difference between the BADGRS
\Mh-\Ms\ scaling and those of other local samples.

\begin{figure*}
    \includegraphics[width=0.45\textwidth]{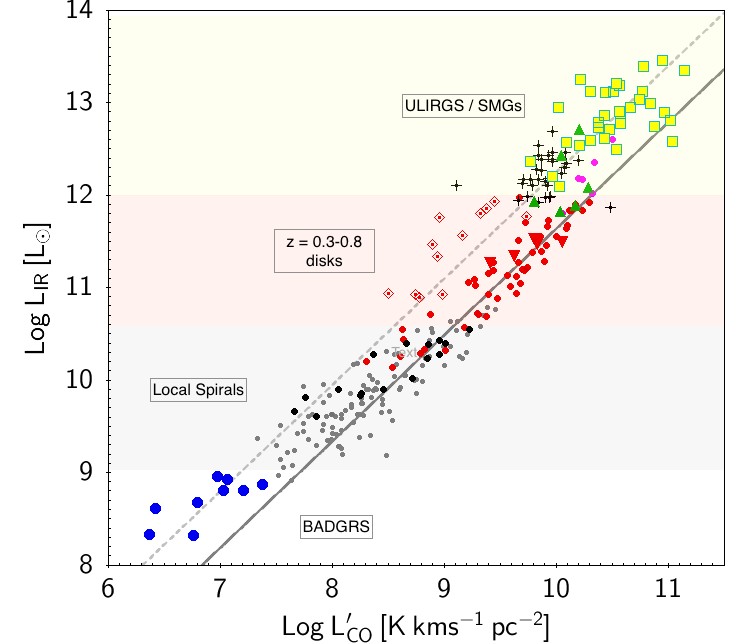}
    \includegraphics[width=0.48\textwidth]{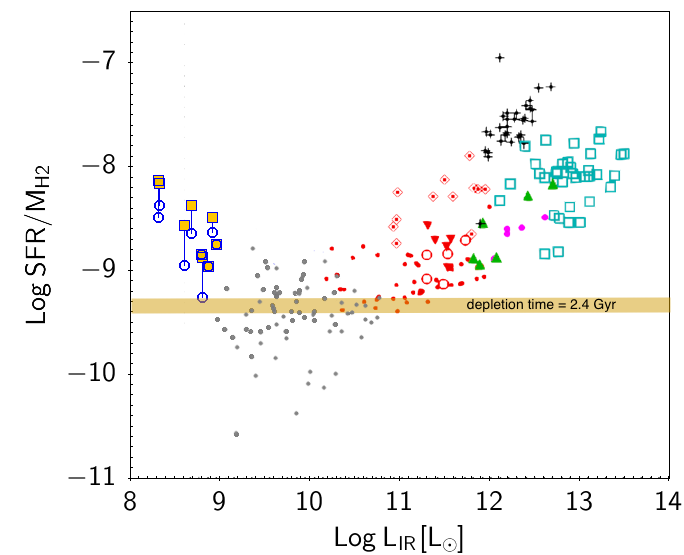}
  \caption{\label{ValesF}{\bf (a)} The correlation between \Lir\ and
    \lcoa for the BADGRS and other samples from the literature. Only
    \COa or \COb measurements are used to avoid uncertainty from
    excitation corrections. Where \COb are used, the conversion used
    in the literature reference is given below. Grey circles: local
    spirals from HRS \citep{Boselli2014}, only late-type HRS sources
    with CO observations covering more than 10 percent of the optical
    disk and which have metallicity measurements are shown. Black
    circles: local disks (\citealt{Wilson2009} and $r_{21}=0.8$
      \citealt{Leroy2008,Leroy2009}); red-- intermediate redshift
      spirals (VALES $z=0.1-0.3$ \citealt{Villanueva2017}, diamonds:
      low SNR CO detections), red triangles: $z\sim 0.4$ disks
      \citep{Geach2011}; black crosses-- local ULIRGs
      \citep{Solomon1997}; green triangles-- $z=0.3-0.8$ 250-\mic\
      selected ULIRGs ($r_{21}=0.75$ \citealt{Magdis2014}); magenta--
      BzK ($r_{21}=0.84$ \citealt{Daddi2010}); yellow squares-- SMG
      (\citealt{Ivison2011,Aravena2016}, $r_{21}=0.8$
      \citealt{Frayer2008}). The dotted and solid lines are the
      relationships derived by \citet{Genzel2010} for mergers and
      normal galaxies respectively. {\bf (b)} Star formation
      efficiency (SFR/\Mh) as a function of \Lir\ using $\aco=0.8$ for
      local ULIRGs \citep{Downes1998,Solomon1997}, \aco(Z) from \citet{Wolfire2010} for the HRS, and
      \aco\ as derived by the authors for the SMG from
      \citet{Ivison2011}. All other points use $\aco=4.3$. For BADGRS
      we use \aco(Z) from Table~\ref{DustT} (blue open circles)
      connected to the values using $\aco=4.3$ (yellow). The brown
      shaded region is the average $\tau_{dep}$ found for local spiral
      disks by \citet{Schruba2011,Bigiel2011}.}

\end{figure*}

The well studied correlation of \Lir\ and \Lco\ for a range
of sources at low and high redshift is shown in
Figure~\ref{ValesF}(a). The relationships for normal star forming
galaxies and `mergers/starbursts' derived by \citet{Genzel2010} are
over-plotted as solid and dashed lines, although there is contention
in the literature as to whether there really is a separate
relationship for starburst mergers and normal galaxies
\citep[e.g.][]{Casey2014,Villanueva2017,Lee2017}. BADGRS, local ULIRGS
and high-z SMGs have a higher \Lir/\Lco\ on average than normal star
forming galaxies at a range of redshifts. Notably the sources with low
CO SNR (open red diamonds) in the VALES sample of 160-\mic\ selected
galaxies \citep{Villanueva2017} also lie in this region of parameter
space. These may either be examples of lower luminosity systems with
high dense gas fractions (like SMGs and ULIRGs), or higher luminosity,
higher redshift examples of BADGRS.

Figure~\ref{ValesF}(b) shows the star formation efficiency (SFR/\Mh)
(or inverse of the depletion timescale) as a function of \Lir\ for the
same samples.  Only low-J \COa\ and \COb\ lines are used to create this
plot meaning that we are not affected by uncertainties in CO
excitation. For $\rm{Log \Lir>11.0}$, the SFR is estimated using the
\Lir-SFR relationship of \citet{Kennicutt1998} for a
\citet{Chabrier2003} IMF, while for the local samples of less obscured
galaxies, a combination of IR, H$\alpha$ and UV emission is used to
capture the total SFR. For BADGRS we have used the FUV and IR emission
in the CO regions to estimate the local SFR as described in
Section~\ref{SampleS}. BADGRS are shown as blue circles using the
metallicity dependent \aco\ from Table~\ref{DustT}, connected to
yellow squares for the values using the Milky Way \aco. Literature samples
have \aco\ as described in the caption.\footnote{There remains an
  active debate over the most appropriate \aco\ value to use for
  higher redshift systems and SMGs, with recent work finding values
  intermediate between the ULIRG and Milky Way values
  \citep{Weiss2005,Ivison2011,Carleton2017,Bothwell2017} but see also
  \citet{Aravena2016}. If we instead used the $\aco=0.8$ ULIRG value
  for the SMGs at high redshift, this shifts the open cyan squares up
  to lie closer to the ULIRGS but it does not affect the discussion
  related to the BADGRS.}


There is a broad correlation, with the most luminous IR systems having
a higher star formation efficiency (or shorter depletion
timescale). Figure~\ref{ValesF}(b) implies that BADGRS have higher
star formation efficiencies (shorter depletion timescales) than other
local galaxies, and in fact have star formation efficiencies similar
to intermediate redshift LIRGS and ULIRGS. There is building evidence
that local low mass, low metallicity galaxies really do have shorter
depletion timescales
\citep{Saintonge2011,Bothwell2014,Huang2014,Hunt2015,Bothwell2016,Amorin2016}.
Such findings were previously dismissed as the result of not
correcting \aco\ for the effects of metallicity variations
\citep{Leroy2011}, but later studies which are able to make these
corrections still find the same trends.

\section{A most unusual molecular ISM in BADGRs}
\label{DiscS}

While BADGRS are very rich in atomic gas and dust, they are
inexplicably deficient in CO, our single dish observations revealing
weak, narrow CO lines with an intensity far lower than that
expected from their dust emission. It is important to understand the
reasons for this, given that the use of both CO and dust
\citep[e.g.][]{Genzel2015,Scoville2017} to estimate molecular gas masses
is commonplace at high redshift. If either tracer can depart from
its usual relationship in a class of galaxy, then clearly it would
lead to systematic biases in the estimation of $\rm{\rho_{H2}}$ and its
evolution.

We now consider some possible explanations for the unusual
dust-$\rm{H_2}$ or dust-CO ratios. In S~\ref{pressureS} we look at the
possibility that conditions for $\rm{H_2}$ are unfavourable and the
ISM is simply H{\sc i} dominated. In S~\ref{attenuationS} we
investigate whether the CO is adequately shielded from
photodissociation by measuring the dust attenuation in the sources in
a variety of ways. In S~\ref{kappaS} we return to the observation that
BADGRS have very cold diffuse dust temperatures and discuss the
implications for the dust properties and geometry, which may have
a bearing on the dust to CO ratio. In S~\ref{CRS} we highlight recent
work which suggests that cosmic rays may be a culprit for destroying
CO throughout the volume of a clouds and finally in S~\ref{burstS} we
look at evidence for a bursty star formation history in BADGRS and
speculate on how that might affect the relative abundance of dust and
CO.

\subsection{Unfavourable conditions for formation of $\rm{H_2}$}
\label{pressureS}

If conditions in the ISM of BADGRS do not favour the formation of
$\rm{H_2}$, they may be H{\sc i} dominated and the dust simply
traces the atomic component. This is often the case in low
metallicity, diffuse dwarf galaxies (which also have low dust
content). The formation of $\rm{H_2}$ requires certain combinations of
column density, pressure and radiation field
\citep[e.g.][]{ElmegreenH21993,PPP2002} and the relationship between
the hydrostatic mid-plane pressure and the molecular fraction
\citep{BR2006,Leroy2008,Bekki2014,Valdivia2016} has a functional form
in gas rich galaxies \citep{Feldmann2012}:

\begin{equation}\label{rmolE}
R_{\rm{mol}} = \Sigma_{\rm{H2}}/\Sigma_{\rm{HI}} = \left(\frac{P_{\rm{ext}}/k}{P_0/k}\right)^{0.7-1.3}
\end{equation}

where the exponent depends on whether the galaxy is in the $H_2$ or
H{\sc i} dominated regime.\footnote{The normalisation and slope of the
  $R_{\rm{mol}}-P_{\rm{ext}}$ relation in \citet{Feldmann2012} were
  found to scale with metallicity and we have included this scaling in
  our determination of $\rm{R_{mol}}$ by using the appropriate slope and normalisation
  from their Fig. 3.} This is related to the molecular fraction
per unit gas mass as $\rm{f_{H2} = (1+1/R_{mol})^{-1}}$.

The hydrostatic mid-plane pressure for a disk which can be dominated by
gas rather than stars (as is the case here) is given by
\citet{Elmegreen1989,Swinbank2011}

\begin{equation}\label{p/kE}
\frac{P_{ext}}{k} = 33.3 \times \Sigma_g \left[\Sigma_g + \left(\frac{\sigma_g}{\sigma_{\ast}}\right)\Sigma_{\ast}\right]
\end{equation}

where $\Sigma_g/\Sigma_{\ast}$ are the gas/stellar surface densities
in units of $\msun \rm{\,pc^{-2}}$, and $\sigma_g$ and $\sigma_{\ast}$
are the gas and stellar velocity dispersions. While we do not yet have
the high resolution H{\sc i} data to hand to directly measure
$\Sigma_{\rm{HI}}$ over the same area as the dust and CO measurements, we can
make a rough estimate of the total gas mass surface density, $\Sigma_g
= \Sigma_{HI} + \Sigma_{H2}$, using the measured dust mass surface
density and the relation from \citet{Draine2014}:

\begin{equation}\label{gdE}
\Sigma_{g,0} = (0.0065 \,\, Z/Z_{\odot})^{-1}\,\Sigma_{d}\times \rm{cos}\,i 
\end{equation}

where we have reduced the constant in the \citet{Draine2014} equation
by a factor 1.4 to be consistent with the value of dust opacity we are
using \citep{Planck2011xix}.

$\Sigma_{\ast,0}$ is the face-on stellar mass surface density averaged
over the \COa\ beam using smoothed stellar mass maps created from
the {\em g} and {\em i} SDSS images and the \citet{Zibetti2009} stellar
mass prescription. The dust, gas and stellar surface
densities are listed in Table~\ref{pressureT}.

We do not yet have direct measures of the stellar and gas velocity
dispersions ($\sigma_g/\sigma_{\ast}$) although in future they will be
obtained from higher resolution H{\sc i} data and optical IFU data.
For simplicity, the pressure calculation will assume that
$\sigma_g/\sigma_{\ast} \sim 1$ as is found for dwarfs in
\citet{Obrien2010}. Alternatively, taking a likely gas velocity dispersion within
$r_{25}$ of $\sigma_g = 15 \kms$ \citet{Tamburro2009} and
$\sigma_{\ast} \sim 25 \kms$ based on measurements from \citet{Martinsson2013}, and
using $V_{rot}/\sigma \sim 0.2$ for late type low mass galaxies does
not affect the conclusions.\footnote{$V_{rot}$ is estimated from the
  H{\sc i} line-widths and corrected for inclination.}

\begin{table*}
\caption{\label{pressureT} CO and dust based gas surface densities and ISM pressure.}
\begin{tabular}{lcccccccccc}
\hline
Pointing & $\rm{\Omega_{beam}}$ & $i$ & $\rm{\Sigma_{H2,CO}}$ & $\Sigma_{d,0}$ & G/D & $\Sigma_{g,0}$ & $\Sigma_{\ast,0}$ & $P_{\rm{ext}}/k$ & $f_{\rm{H2,th}}$ & $\frac{f_{\rm{H2,th}}}{f_{\rm{H2,CO}}}$ \\
     &  ($\rm{kpc^2}$) &  & ($\rm{\msun\, pc^{-2}}$) & ($\rm{\msun\, pc^{-2}}$)&  & ($\rm{\msun\, pc^{-2}}$) & ($\rm{\msun\, pc^{-2}}$) & ($10^4\,\rm{K\,cm^{-3}}$) &  &  \\ 
\hline
N5584 C & 11.3 & 56$^1$ & 5.1 & 0.21 & 154  & 32.3  & 76.2 & 11.7   & 0.8 & 5     \\
N5584 O1 & ..  & .. & 3.6 & 0.23 & 228  & 52.3  & 8.8  & 10.6   & 0.7 & 10   \\
N5584 O2 & ..  & .. & 3.6 & 0.18 & 173  & 31.1  & 17.0 & 5.0    & 0.6 & 5   \\
\hline
N5496 C &  9.3 & 79$^2$ & 1.0 & 0.14 & 177  & 24.2  & 22.7 & 3.8    & 0.6 & 15   \\
N5496 O1  & .. & .. & 1.4 & 0.11 & 344  & 37.8  & 9.7  & 6.0    & 0.5 & 13   \\
N5496 O2  & .. & .. & 1.2 & 0.05 & 361  & 16.9  & 4.5  & 1.2    & 0.2 & 3    \\
\hline
U9215 C  & 8.1 &  54$^1$ & 5.7 & 0.34 & 274 & 93.0  & 45.3 & 42.8   & 0.8 & 14   \\
U9215 O1  & .. & ..  & 2.0 & 0.15 & 329 & 49.3  & 11.9 & 10.0   & 0.6 & 14 \\
\hline        
U9299 & 9.8     & 61$^1$ & 0.5 & 0.09 & 255 &  23.0  & 15.9 & 3.0   & 0.4 & 22 \\
\hline
\end{tabular}
\flushleft{\small $\rm{\Omega_{beam}}$: beam area, the physical region sampled by the
  115-GHz beam and used to derive the surface densities. {\em i}:
  inclination, used to correct measured values to
  face-on. $\Sigma_{H2,CO}$: CO inferred $H_2$ surface density
  corrected to face-on. $\Sigma_{d,0}$: dust mass surface density
  measured in the area of the IRAM \COa beam and corrected to
  face-on. G/D: gas/dust ratio expected from the metallicity according
  to \citet{Draine2014} and Eqn~\ref{gdE}. $\Sigma_{g,0}$: Total
  face-on H surface density derived from $\Sigma_{d,0}$ and
  G/D. $\Sigma_{\ast,0}$: face-on stellar mass surface density
  averaged in the region of the \COa beam using stellar mass maps
  created from the {\em g} and {\em i} SDSS images and the
  \citet{Zibetti2009} stellar mass prescription. $P_{\rm{ext}}/k$:
  mid-plane hydrostatic pressure using Eqn~\ref{p/kE}.
  $f_{\rm{H2,th}}$: molecular fraction $\rm{\Mh/(M_{HI}+\Mh)}$ from
  Eqn~\ref{rmolE}. The final column is the ratio of $f_{\rm{H2}}$
  estimated theoretically based on the dust density and
  Eqn~\ref{rmolE} and that estimated directly from the CO
  measurements. References: $^1$
  LEDA\footnote{http://leda.univ-lyon1.fr/} \citep{Makarov2014}, $^2$
  \citet{Guthrie1992}.}
\end{table*}

The pressures we estimate in 2-3 kpc regions using the above
assumptions are in the range $\rm{P_{ext}/k \sim 1-40\times
  10^4\,K\,cm^{-3}}$. At these pressures we expect $\Sigma_{\rm{H2}}\geq
\Sigma_{\rm{HI}}$ at 7/9 locations, the exceptions being
the outer disk of NGC~5496 and UGC~9299. In the final column of
Table~\ref{pressureT} we compare the molecular fraction derived using
the arguments just described ($\rm{f_{H2,th}}$), to that estimated
directly from CO using the metallicity dependent \aco(Z):
\[
\rm{f_{H2,CO}= \Sigma_{\rm{H2,CO}}/\Sigma_g}
\] 
The surface density of molecular gas as traced by CO is a factor 3--22
(mean 11) times lower than expected given the dust emission, metallicity and stellar mass
surface density measured in each region.

Resolved studies of nearby spirals and dwarfs
\citep{Bigiel2008,Leroy2008} find a marked transition from H{\sc
  i}--$\rm{H_2}$ with equal contributions occurring at a gas mass
surface density of $\sim 10 \msun \rm{pc^{-2}}$. Our dust based
$\Sigma_g$ estimates in Table~\ref{pressureT} are all at or above this
value, although the transition threshold will increase with decreasing
metallicity.

In summary, it appears that lack of favourable conditions for
formation of $\rm{H_2}$ in the ISM of these galaxies is not the
culprit, but {\em this assumes that the dust is a reliable tracer of
  the total gas, modulo a metallicity dependence}. A future analysis
of high resolution H{\sc i} imaging will allow us to more fully
examine the balance between dust, atomic and molecular gas within the
disks and star forming regions of these sources.

\subsection{Photodissociation of CO due to lack of shielding}
\label{attenuationS}

The usual reason that the ability of CO to trace $\rm{H_2}$ breaks
down is a lack of shielding for the fragile CO molecule from UV
photo-dissociation \citep[e.g.][]{Lee2015}. When the metallicity (or
dust/gas ratio) is low, the region of a molecular cloud at which
$\rm{A_v}$ is high enough to shield the CO is very much reduced
\citep{Wolfire2010,Glover2011,PPPP2010}. This in turn means that the
CO is only present in the very central cores of the dense clouds, and
can be missing from the cloud envelopes or from the entirety of less
dense clouds. However, at the metallicities of the BADGRS
($0.5-1\,\rm{Z_{\odot}}$) CO is believed to trace $\rm{H_2}$ to within
a factor of 2 of the Milky Way \aco\ scaling \citep{Bolatto2013}. Below these metallicities, CO is efficiently
photo-dissociated leading to regions of `CO dark' molecular gas where
\aco\ can be 10-100 times larger than the Milky Way value \citep{Wolfire2010}. Evidence for CO dark
gas has been found in low metallicity dwarf galaxies
\citep[e.g.][]{Leroy2011,Shi2014}, and in low $\rm{A_v}$ regions of the
Milky Way \citep{Planck2011xix}. A simple comparison of the \Mh/\Md\
ratios for BADGRS (7--27) with those expected for galaxies of similar metallicity (100--150)
(e.g. Fig~\ref{GrossiF}) would suggest that 70-90 \% of the molecular
gas in the BADGRS is `CO dark' and not recovered even with a
metallicity dependent \aco.

\begin{figure}
\includegraphics[width=0.5\textwidth]{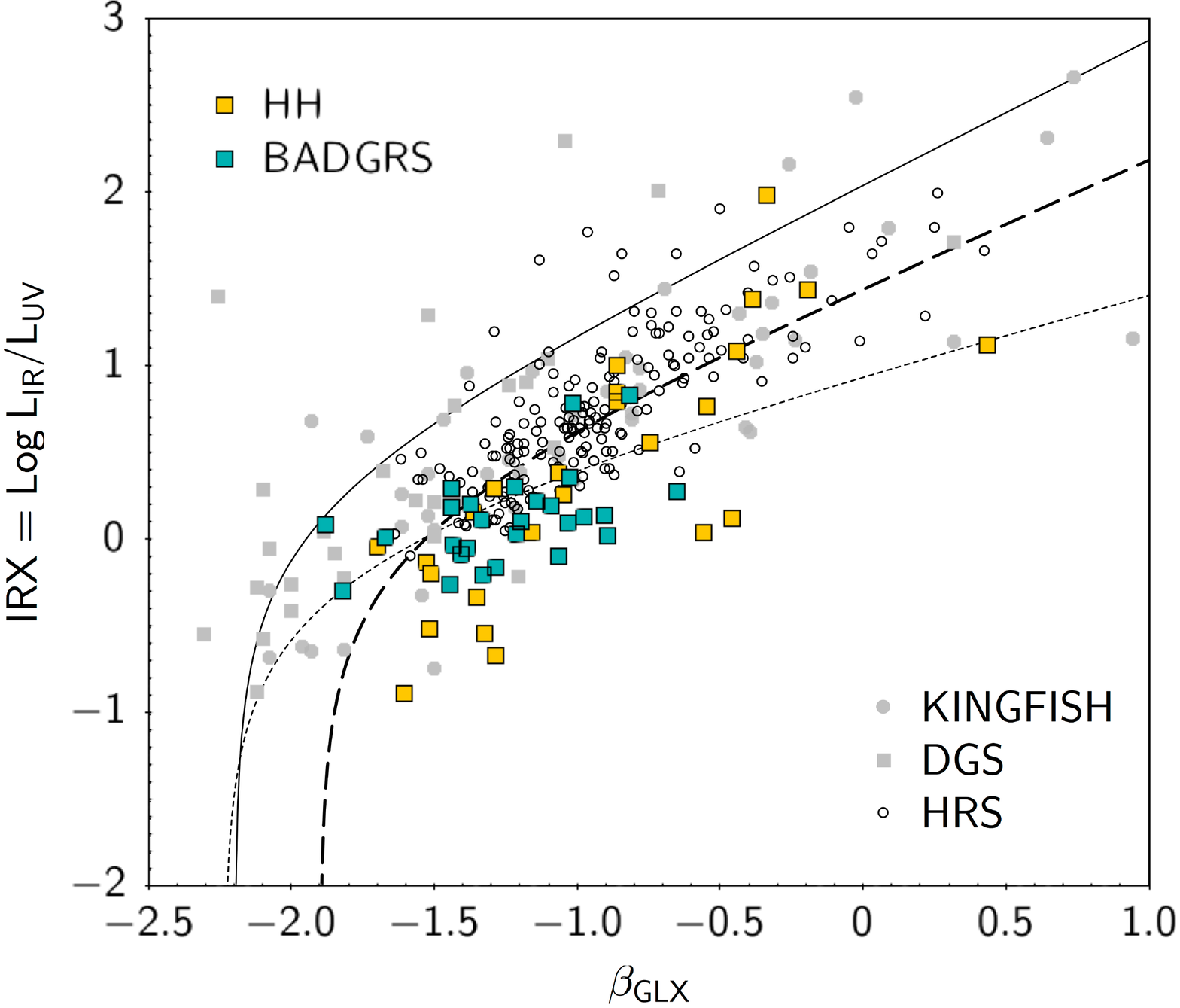}
\caption{\label{irxF} IRX-$\beta$ for star forming galaxies in the
  HRS, KINGFISH and DGS samples (grey and black points)
  \citep{Ciesla2014,deVis2016,Skibba2011,Madden2013,Faisst2017}. Star
  forming galaxies from the dust (HAPLESS) and H{\sc i} (HIGH)
  selected samples are shown as coloured squares. Those designated as
  BADGRS based on FUV-K colour and dust/stellar mass ratio are blue
  while those which are redder and/or more dust poor are gold.
  Attenuation curves for for local starbursts \citep{Overzier2011},
  normal galaxies \citep{Takeuchi2012} and the SMC \citep{Bouwens2016}
  are shown as solid, dashed and dotted lines respectively. The BADGRS
  lie offset in the IRX-$\beta$ space compared with the other samples
  (including the DGS), which tend to have bluer UV colours at the same
  attenuation. However so also do many of the H{\sc i} selected HIGH
  sample of \citet{deVis2017}.}
\end{figure}


Given that these sources were selected on the basis of their dust
emission in a volume limited survey, it would be rather surprising for
them to be devoid of CO due to a lack of dust shielding. Estimates of
$\rm{A_{FUV}}$ for HAPLESS, HIGH and the HRS were made by
\citet{deVis2017} using the ratio of the attenuated and intrinsic FUV
luminosities from {\sc magphys} fitting\footnote{This is consistent
  with the $\rm{A_{FUV}}$ we would derive using the empirical
  relationship from \citet{Cortese2008} for a similar range of Hubble
  types, and to the relationship between IRX and $\rm{A_{FUV}}$ from
  \citet{Hao2011}.} Interestingly, BADGRS have much lower attenuation
($\rm{A_{FUV}}\sim0.2-0.8$) than normal spirals ($\rm{A_{FUV}}=1-3$),
even at the same dust mass.

{\em Why do these galaxies have such low attenuation despite their
  dust and metal content being relatively high?\/} That a given
quantity of dust can produce differing amounts of attenuation is
directly a question of dust geometry and/or grain properties which
affect the attenuation curve. Changes in these may lower the effective
shielding for CO at a given metallicity (or dust/gas ratio).



The infrared/UV luminosity ratio, IRX, is a calorific measure of the
UV attenuation which is insensitive to geometry (though it does
require knowledge of how much dust is heated by UV versus optical
photons, \citealt{Cortese2008,Viaene2016}). The relationship between
IRX and UV slope ($\beta$) usefully {\em is} affected by the dust
attenuation curve (which is a combination of star/dust geometry and
intrinsic dust absorption and scatterting properties), for more
details see e.g.
\citet{Gordon2000,Kong2004,Boquien2009,Buat2012,Mao2014,Battisti2016,Faisst2017}.
The UV slope is also sensitive to the age of the stellar population
which dominates the UV light, and with complex star formation
histories there are many degeneracies, \citep[e.g.][]{Kong2004,Boquien2012,Mao2012}.

The IRX-$\beta$ relation for galaxies with type later than S0 and $\rm
Log\,sSFR > -11.0$ from the combined HAPLESS and HIGH (HH), HRS, KINGFISH and DGS samples
is shown in Figure~\ref{irxF}. These samples have good quality {\em
  Herschel} data meaning that \Lir\ is well determined (earlier studies
have often relied on templates to infer \Lir\ from data on the short
wavelength side of the dust SED which makes them less reliable to
compare to). Also shown are the relations for (solid) starbursts 
\citep{Overzier2011}, (dashed) normal star forming galaxies
\citep{Takeuchi2012} and (dotted) the SMC \citep{Bouwens2016}, which has a
steeper extinction curve. Most of the BADGRS lie on or below the SMC dust
curve, having lower UV attenuation at the same UV colour (or redder UV
colours at the same UV attenuation). To summarise the differences, the
average fraction of HRS, KINGFISH and DGS galaxies lying below all
three attenuation curves is 15, 15 and 3 percent respectively. For the combined 
HAPLESS and HIGH samples the fraction is 63 percent, while 74
percent of BADGRS lie below the SMC curve.

Several papers have claimed that high redshift ($z>5$) galaxies lie
below the typical starburst and spiral relations using the Calzetti
dust law \citep{Siana2009,Reddy2010,Bouwens2016,Faisst2017}. A more
detailed investigation into the IRX-$\beta$ relationship for the
BADGRS (and HAPLESS and HIGH in general) will be presented elsewhere
(Dunne et al. in prep), where we will explore how much of the offset
can be ascribed to dust properties and how much to differences in star
formation history, and investigate the possibility of using BADGRS as
low redshift ISM analogues for high redshift sources.

\subsection{Dust emissivity and/or geometry effects}
\label{kappaS}

One of the other unusual aspects of the BADGRS ISM is that the dust in
the diffuse ISM is much colder (12--16 K) than in spirals or low
metallicity dwarfs ($\sim 18-31$ K;
\citealt{Remy-Ruyer2015,Clark2015,Hunt2015}). The temperature of dust grains
in thermal equilibrium with the interstellar radiation field can be
written as:

\[
T_d^{4+\beta} \propto 4\pi \int^{3\mu m}_{0.09} J_{\lambda} \sigma_{\lambda} d\lambda
\]

where $4\pi J_{\ast}$ is the integrated intensity of the radiation
field, $\sigma_{\lambda}$ is the dust absorption cross section (which
is a function of wavelength) and $\beta$ is the dust emissivity
index. For similar dust properties (i.e. similar $\sigma_{\lambda}$),
such a temperature difference would require an ISRF $20-120\times$
lower in BADGRS than that in spirals and low metallicity dwarfs. Using
the relationship between ISRF and dust temperature from
\citet{Hunt2015} (which is based on the \citet{Draine2007} model
opacities), we estimate that the BADGRS should have a $U_{\rm{min}}$
(the radiation field heating the bulk of the dust mass) of $\sim
0.05-0.1$ times the local solar neighbourhood ISRF, a factor 10--20
lower than the KINGFISH average \citep{Hunt2015}. While BADGRS have
low stellar mass surface densities, their extreme blue SED means that
their FUV surface brightness is much higher than typical spirals. By
integrating the UV--optical SED fitted with {\sc magphys} by
\citet{deVis2017} and dividing by the area of the aperture used to
measure the emission, we can directly measure the bolometric surface
brightness of radiation finding values of $\chi =
1.7-2.3\times10^{-5}\,\rm{W\,m^{-2}}$ (not corrected for internal
extinction). These values are similar to the local solar neighbourhood
ISRF of $\chi_{\odot} = 2.2\times 10^{-5}\, \rm{W\,m^{-2}}$ determined
by \citet{Mathis1983}. Thus the global averaged radiation field is not
compatible with the equilibrium dust temperature for dust emissivities
similar to those used by \citet{Draine2007}, or for dust in the Milky
Way.


Possible solutions to this conundrum could be (A) the dust is mainly
in dense clumps which are shielded from the ISRF, whilst remaining
optically thin to FIR/sub-mm emission, or (B) the dust is less
efficient at absorbing in the UV/optical and/or more efficient at
radiating in the FIR/sub-mm. This would translate into a dust
absorption coefficient, $\sigma_{\lambda}$, a factor $\sim 10-20$
different (lower in the UV/optical or higher in the sub-mm) in BADGRS
than in the Milky Way and other nearby spirals. Either scenario would
allow the dust to remain at a low temperature despite the UV radiation
fields, since in case (A) most of the dust does not see the UV photons
and in Case (B) the dust does not absorb them as easily; both also
naturally explain the low UV attenuation.

The {\em Planck} mission found cold clumps in the Milky Way with
$T_d\sim $13 K and an average density of $2\times10^3\,\rm{cm^{-3}}$
\citep{Planck2011xxii,Juvela2017}. If the dust in BADGRS were located
in similarly dense clumps then one might expect prolific CO emission
since clearly the clumps are well shielded from the UV field. However,
CO also freezes out onto dust grains in such conditions
\citep{Fontani2012} and this may well account for a reduction in the
CO emission relative to that at sub-mm wavelengths. A geometrical
explanation such as this could be investigated using radiative
transfer methods, we defer this to a future paper.

If, on the other hand, the dust opacity were different by a factor
10--20, the lack of CO can then be explained by the lack of shielding
provided by the dust. If the dust is a much poorer absorber then the
UV photons can penetrate deeper into the molecular clouds and destroy
the CO, while alternatively if the dust is a more efficient emitter in
the sub-mm, we will have overestimated the mass of dust using the
Milky Way sub-mm opacity $\kappa_{250}$. The dust opacity is
determined by the composition and structure of the grains, so such a
dramatic change in dust properties across an entire galaxy would be
extremely unusual and difficult to explain. The emissivity of dust at
sub-mm wavelengths has been observed
\citep{Bot2009,Planck2011xix,Martin2012,Planck2014xi,Ysard2015,Fanciullo2015,Planck2016xxix},
to increase by a factor $\sim 2$ in dense environments within the
Milky Way and varies within the diffuse ISM by smaller factors. It is
known that the conditions in dense clouds can lead to changes in the
grain structure via aggregation and the formation of ice mantles
\citep{Kohler2012} but such mechanisms do not have time to operate in
the diffuse ISM. Instead it is proposed that some other processing,
such as FUV irradiation and changes in mantle composition and
thickness in the \citet{Jones2013} dust model, as investigated by
\citet{Ysard2015} and \citet{Fanciullo2015}, can serve to produce (at
least qualitatively) the changes measured in the dust SED of the high
latitude Milky Way by \citet{Planck2016xxix}. It must be noted though
that the changes in opacity required in the Milky Way are only of
order 20 percent (40-50 percent for the extremes). To explain the dust
temperatures and CO emission in BADGRS, we would need to invoke far
more radical changes.

We defer further investigation of dust properties and geometry for
when we have higher resolution sub-mm data and optical measures of
attenuation across the galaxies from IFU data.

\subsection{Depletion of CO by cosmic rays}
\label{CRS}

Cosmic rays from recent star formation can dissociate CO throughout
the volume of a molecular cloud, not just at its surface as is the
case in PDRs. \citet{Bisbas2015,Bisbas2017} show that for certain
conditions of gas density and cosmic ray ionisation rate ($\zeta_{\rm CR}$),
CO may be almost completely destroyed rendering only the densest parts
of the clouds visible. The effect of cosmic rays are not confined to
extreme star formation environments, and even at $\zeta_{\rm CR}$ of 10
times the MW value, there is a marked effect on the CO abundance
relative to $\rm{H_2}$. Clouds that are low density are more
susceptible, so if the molecular clouds in BADGRS were more diffuse
than typical GMCs (e.g. $n_{\rm H2}\sim 100-250\rm{cm^{-3}}$), they could
become CO depleted even at $\zeta_{\rm CR}\sim \zeta_{\rm MW}$. This would
necessitate a population of diffuse, high column density $\rm{H_2}$
clouds, which may not be gravitationally bound. A more detailed study
will be possible with high resolution synchrotron data, which will
reveal the local $\zeta_{\rm CR}$ across the galaxies, and ALMA data to
determine the properties of the molecular gas via high resolution CO
and dust observations.

\subsection{Temporary changes in the CO/dust ratio.}
\label{burstS}

\begin{figure}
\includegraphics[width=0.45\textwidth]{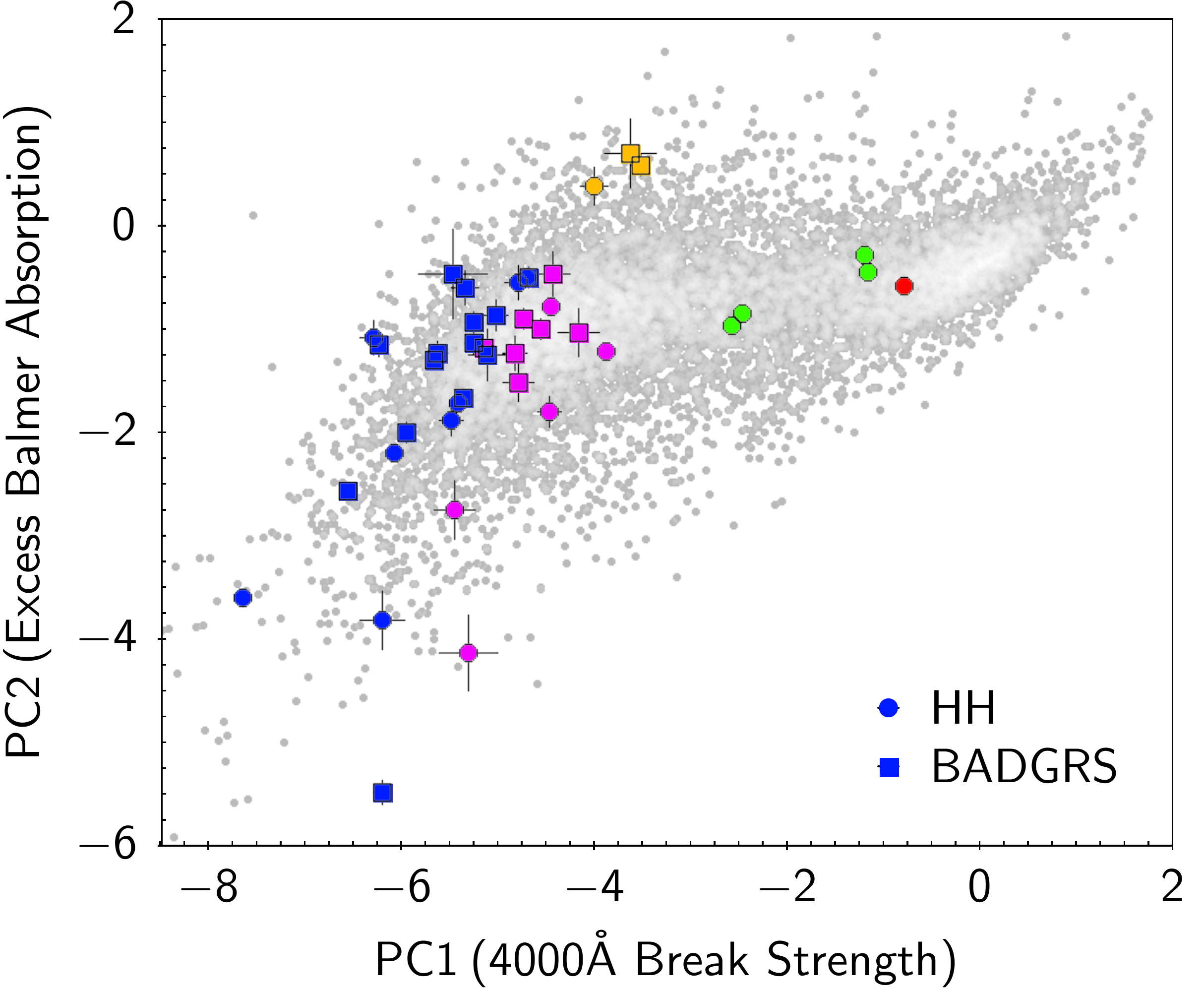}
\caption{\label{SFHF} The spectral diagnostic of star-formation
  history from \citet{Wild2007} showing the location of BADGRS and
  non-BADGRS within the combined HAPLESS and HIGH samples (HH). The
  background points are those GAMA galaxies with good quality spectra
  in the range $0.05<z<0.11$ and $\rm{log\Ms>9.0}$. The colours of the
  points indicate their spectral classifications: blue (starbursts),
  purple (star forming), yellow (post-starburst), green (green
  valley), red (quiescent). The spectra support the broad-band SED
  results from {\sc magphys} indicating that most BADGRS are currently
  or have recently experienced bursts of star formation. The four
  pilot galaxies are all in the starburst category.}
\end{figure}

A range of evidence indicates that many BADGRS are undergoing, or have
recently experienced, bursts of star formation\footnote{Not to be
  confused with a `starburst' classification based on a threshold in
  the strength of the current SF relative to past SF. Here we simply
  mean an increase in the recent SF rate compared to the past
  average}. Their UV colours (uncorrected for attenuation) indicate a
timescale of $t_{\rm{burst}}<300$ Myr \citep{Bianchi2005}, while an
exploration of the star formation histories from {\sc magphys}
indicates a higher probability of a burst in the last few hundred Myr
for BADGRS compared to HRS. The optical spectra of the BADGRS show
strong emission lines and a small 4000\AA\, break consistent with
ongoing star formation. A classification of the optical spectra of the
combined dust and H{\sc i} selected samples (HH) based on a Principle
Component analysis of the stellar continuum in the 4000\AA\, break
region is shown in Fig.~\ref{SFHF} (following
\citealt{Wild2007}). BADGRS reside mostly in, or on the border of, the
starburst region in Figure~\ref{SFHF}, with 70 percent either
currently or having recently experienced a starburst (compared to 47
percent of the non-BADGR members of the HH sample). The underlying
density of points are from the optically selected GAMA survey
\citep{Hopkins2013} and show the typical bimodality of optically
selected samples. This illustrates how rare the starburst and
post-starbursts are in a typical optical sample, in contrast to their prevalence
in our ISM selected samples.

\begin{figure}
\includegraphics[width=0.47\textwidth]{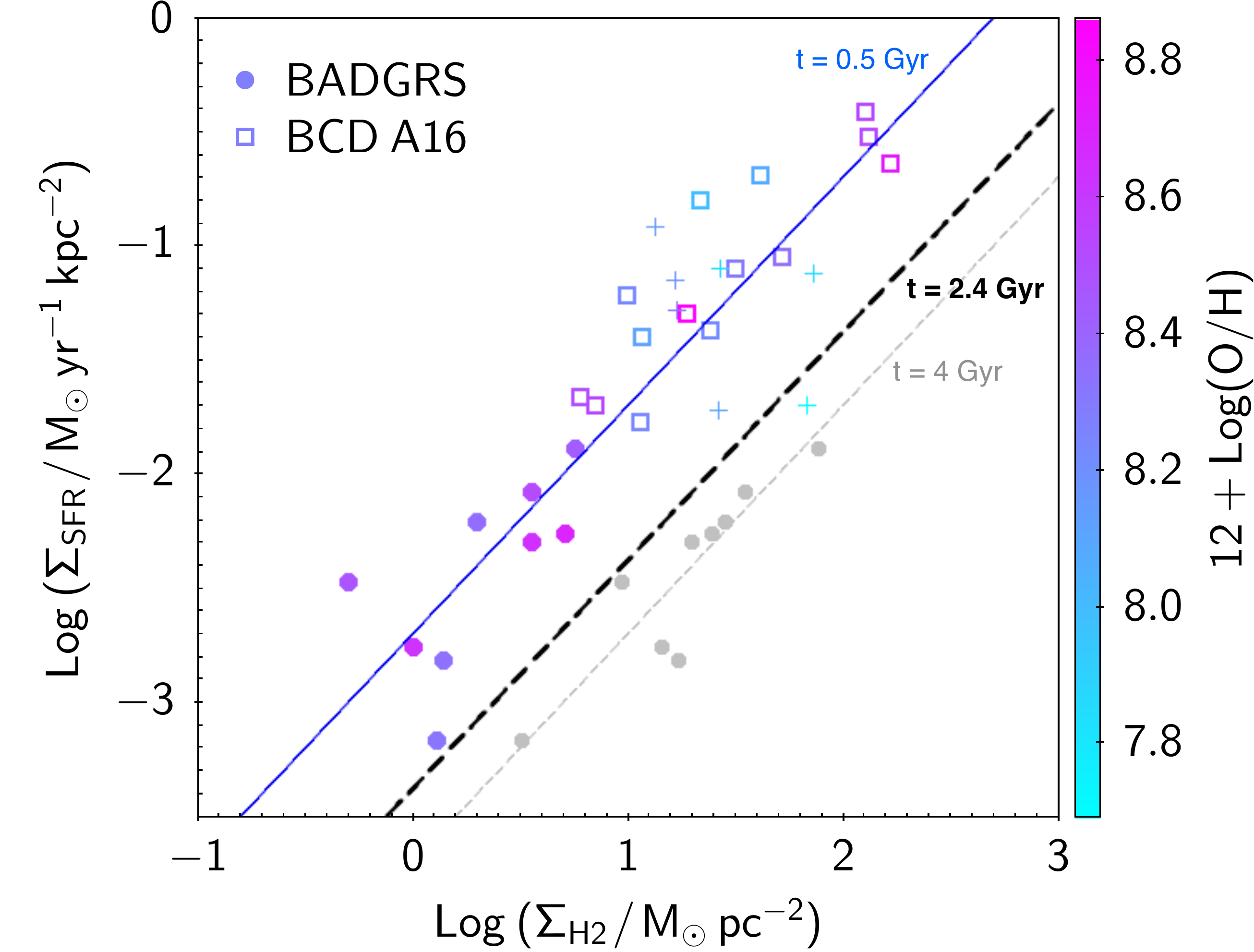}
\caption{\label{KSF} Star formation rate surface density versus
  molecular gas surface density for the BADGRS (solid circles) and BCD
  galaxies from \citet{Amorin2016} (open squares, with crosses as
  upper limits in $\rm{\Sigma_{H2}}$). Both BADGRS and BCD samples use
  similar methods to correct \aco\ for metallicity. Metallicity is
  shown as a colour scale, revealing that the BCD non-detections are
  the lowest metallicity sources. BADGRS are more diffuse objects than
  BCD, but have similarly short molecular depletion timescales
  compared to samples of local spiral galaxies ($\rm{\tau_{dep}\sim2.4}$ Gyr; thick
  black dashed line). The dust based $\rm{\Sigma_{H2}}$ estimates for BADGRS
  (as described in the text) are the grey circles which lie closer
  to the average values found for more massive, lower sSFR
  disks. Depletion timescales of 0.5 Gyr (purple) and 4 Gyr (grey
  dashed) are also shown for reference.}
\end{figure}

As noted in Figure~\ref{ValesF}(b) the BADGRS appear to have higher
SFE (SFR/\Mh) than local spirals. This may also be an indication of a
more bursty star formation mode. Figure~\ref{KSF} shows the surface
densities of star formation rate and molecular gas for BADGRS and a
sample of Blue Compact Dwarf galaxies (BCD) from \citet{Amorin2016},
who use a very similar metallicity dependent \aco. This is the
Kennicutt-Schmidt (K-S) relation \citep{Schmidt1959,Kennicutt1998},
where $\rm{\Sigma_{SFR} = A\times (\Sigma_{H2})^N}$. For the BADGRS we
use $\rm{\Sigma_{SFR_{(UV+IR)}}}$ and $\rm{\Sigma_{H2,CO}}$ derived from
the FUV images, FIR images and CO measurements using \aco(Z)
(Table~\ref{pressureT}). Both the BADGRS and BCDs have higher SFR per
unit $\rm{H_2}$ surface density compared to the relations for local
spiral disks (black dashed line) which show a constant depletion
timescale of $\rm{\tau_{dep}}\sim 2.4$ Gyr
\citep{Bigiel2008,Schruba2011,Bigiel2011,Leroy2011,Leroy2013}. \footnote{We
  find a similar $\rm{\tau_{dep}}$ for the sample of late type spirals
  from the HRS when using the \citet{Wolfire2010} metallicity
  dependent \aco\ and the same sample as plotted in
  Fig~\ref{ValesF}(b).}  The BADGRS are generally higher metallicity
than most of the BCDs, and although they have lower surface densities
of both SFR and molecular gas (i.e. they are more diffuse galaxies),
they have a similar star formation law when we derive $\Sigma_{H2}$
from $I_{CO}$. Blue Compact Dwarfs are examples of galaxies undergoing significant
bursts of star formation, hence their high SFE is thought to be
related to this mode of star formation. 

As the tendency to have a recent burst seems prevalent in BADGRS we
hypothesise that a temporary increase in the dust-to-CO (or
dust-to-$\rm{H_2}$) ratio may occur following a burst of star
formation. The mechanism for this may include rapid dust formation in
supernovae
\citep{Dunne2003,Dunne2009,Matsuura2011,Gomez2012,Gall2014,Indebetouw2014,deLooze2017}
and possible temporary depletion of the local $\rm{H_2}$ reservoir and
dispersal of molecular clouds by turbulence. For very recent
starbursts there may be changes in the chemical composition of the ISM
and dust due to the relative abundance of products from massive stars,
enriching the ISM with O and producing a lower C/O ratio than
normal. This may have the effect of reducing CO abundance as well as
increasing O-rich dust species. If correct, this hypothesis implies
that large deviations in the dust-CO or (dust-$\rm{H_2}$) ratio must
happen regularly in the life of intermediate mass, gas-rich galaxies,
since BADGRS make up more than half of the galaxies in the H-ATLAS
local volume survey, HAPLESS. As it is a dust mass selected sample,
HAPLESS would preferentially include galaxies near this hypothetical
`peak' in their dust cycle.

Alternatively, if we use instead the dust emission and theoretical
arguments (Section~\ref{pressureS}, Table~\ref{pressureT}) to estimate
$\rm{\Sigma_{H2}}$ for BADGRS
($\rm{\Sigma_{H2,d}=f_{H2,th}\times\Sigma_g}$), the SFE drops
dramatically and the depletion times become similar to or longer than
the more massive galaxies (grey circles in Fig~\ref{KSF}).

In summary, either BADGRS have an enhanced SFE (short molecular
depletion timescale, $\rm{\tau_{dep}} \sim 0.5$ Gyr) or, they have
normal SFE with $\rm{\tau_{dep}\sim 2.4-4}$ Gyr but some mechanism
  means that they are more deficient in CO relative to their $\rm{H_2}$
  content than the current understanding of metallicity scaling
  relations predict. Measuring the star formation histories across
  these galaxies to determine how much of an influence a bursty
  history might have will be possible using IFU data, to be presented
  in Ballard et al. {\em in prep}.

\section{Conclusions}
\label{conclusions}

We have presented measurements of the $^{12}$CO J=1,2,3 lines for 9
positions in four representative examples of extremely blue but dusty
gas rich galaxies (BADGRS) found in the first dust mass selected
sample from the {\em Herschel}-ATLAS survey.

\begin{itemize}
\item{The CO lines in these sources are very weak with $T_p = 5-30$ mK
  (\tmb), the linewidths are narrow in some cases, and the implied
  $\Sigma_{H2,CO}=0.5-6\,\rm{M_{\odot}\,pc^{-2}}$ is 6--10 times less than
  the interarm density in M51, averaged over a similar area.}
\item{The BADGRS exhibit a range of excitation conditions from their
  low-J CO lines, much the same as other local spirals.}
\item{Using a metallicity dependent \aco, we find \Mh/\Md\ ratios of
  $7-27$, which are a factor $\sim 10$ lower than other local samples
  with consistently estimated metallicities, \Mh\ and dust masses. The
  \Mh/\Md\ ratio is found to be insensitive to metallicity.}
\item{The BADGRS have similar \Mh-\Ms\ properties compared to other
  local samples, although the scatter in these relations is large. }
\item{Using dust as a tracer of total gas surface density, we show
  that the average ISM pressure in the regions where CO is measured
  should be high enough that at least 50 percent of the gas there is
  in molecular form. The $\rm{H_2}$ fractions inferred in this way are
  on average 11 times higher than those indicated by the CO emission.}
\item{The IRX-$\beta_{UV}$ relation for BADGRS is offset to redder UV
  colour for a given attenuation compared to local star forming
  galaxies and starbursts. The majority (74 percent) of BADGRS lie on
  or below an SMC-like dust attenuation law, and in this way are
  similar to $z\sim 5$ galaxies \citep{Faisst2017}.}
\item{The diffuse dust temperatures in the BADGRS are much colder
  (13--16 K) than in typical spirals or low metallicity dust poor
  dwarfs (18--31 K). For the same dust properties, this would require
  an ISRF 10--20 times lower than the local solar neighbourhood value,
  however, the observed radiation surface densities are in fact
  equivalent to the solar neighbourhood value. In order to explain
  this discrepancy either a radically different dust geometry
  (clumpier) or different dust properties (size distribution,
  composition, opacity) compared to the Milky Way and other local
  galaxies is required.}
\item{Using the CO based \Mh, BADGRS are significantly offset from the K-S relation for local
  spiral galaxies, with shorter molecular gas depletion timescales of
  $\sim 0.5$ Gyr, even when using a metallicity dependent \aco. In
  this respect they are similar to low metallicity Blue Compact Dwarf
  galaxies, which have bursty star formation histories. Analysis of
  their SEDs and optical spectra also suggests BADGRS are currently,
  or have recently, experienced bursts of star formation. There may
  be a link between this and a genuinely enhanced
  dust-to-molecular-gas ratio.}
\end{itemize}

Future papers will describe the data from ongoing high resolution
Jansky-VLA and ALMA studies of the atomic and molecular ISM and from
optical IFU imaging of the stellar properties and kinematics. From
this full data set and with radiative transfer modelling of the
galaxies, we will be able to remove some of the assumptions made here
and address these intriguing options in turn to solve the mystery
posed by the BADGRS.

\section*{Acknowledgements} Thanks to A. Faisst, L. Hunt, V. Villaneuva and M. Grossi for their help in providing data points for literature samples, and to S. Schofield for Figure~\ref{SEDstackF}. We thank Y. Mao, N. Bourne, M. Michalowski and S. Viaene for helpful comments. LD and SJM acknowledge support from the European Research Council Advanced Investigator grant, COSMICISM and Consolidator grant, CosmicDust. IO, ZZ and RJI acknowledge support from the European Research Council Advanced Investigator grant, COSMICISM. HLG and PC acknowledge support from the European Research Council Consolidator grant, CosmicDust. 
This work is based on observations carried out under project number
077-14 with the IRAM 30m telescope. IRAM is supported by INSU/CNRS
(France), MPG (Germany) and IGN (Spain)."  The {\it {\em
    Herschel}}-ATLAS is a project with {\it {\em Herschel}}, which is
an ESA space observatory with science instruments provided by
European-led Principal Investigator consortia and with important
participation from NASA. The H-ATLAS web-site is
http://www.h-atlas.org. GAMA is a joint European-Australasian project
based around a spectroscopic campaign using the Anglo- Australian
Telescope. The GAMA input catalogue is based on data taken from the
Sloan Digital Sky Survey and the UKIRT Infrared Deep Sky
Survey. Complementary imaging of the GAMA regions is being obtained by
a number of independent survey programs including GALEX MIS, VST KIDS,
VISTA VIKING, WISE, {\em Herschel}-ATLAS, GMRT and ASKAP providing UV
to radio coverage. GAMA is funded by the STFC (UK), the ARC
(Australia), the AAO, and the participating institutions. The GAMA
website is: http://www.gama-survey.org/. We acknowledge the usage of
the HyperLeda database (http://leda.univ-lyon1.fr).
\bibliographystyle{mnras}
\bibliography{paper_submitted}
\bsp
\label{lastpage}
\end{document}